\documentclass[%
superscriptaddress,
reprint,
amsmath,amssymb,
aps,
pre,
]{revtex4-2}

\usepackage[english]{babel}

\usepackage{graphicx}
\usepackage[percent]{overpic}
\usepackage{dcolumn}
\usepackage{bm}

\usepackage{hyperref}
\usepackage{color}
\definecolor{rltred}{rgb}{0.75,0,0}
\definecolor{rltgreen}{rgb}{0,0.6,0}
\definecolor{rltblue}{rgb}{0.3,0.3,1}

\usepackage{placeins}

\hypersetup{colorlinks,%
	hypertexnames=true,%
	linkcolor=rltred,%
	citecolor=rltgreen,%
	linktocpage=true}

\begin{document}
\title{Non-equilibrium correlation dynamics in the one-dimensional Fermi-Hubbard model: A testbed for the two-particle reduced density matrix theory}

\author{Stefan Donsa}
\affiliation{Institute for Theoretical Physics, Vienna University of Technology,
	Wiedner Hauptstra\ss e 8-10/136, 1040 Vienna, Austria, EU}

\author{Fabian Lackner}
\affiliation{Institute for Theoretical Physics, Vienna University of Technology,
	Wiedner Hauptstra\ss e 8-10/136, 1040 Vienna, Austria, EU}

\author{Joachim Burgd\"orfer}
\affiliation{Institute for Theoretical Physics, Vienna University of Technology,
    Wiedner Hauptstra\ss e 8-10/136, 1040 Vienna, Austria, EU}

\author{Michael Bonitz}
\affiliation{Institut f\"ur Theoretische Physik und Astrophysik, Christian-Albrechts-Universit\"at zu Kiel, D-24098 Kiel, Germany, EU}
\affiliation{Kiel Nano, Surface and Interface Science KiNSIS, Kiel University, Germany, EU}

\author{Benedikt Kloss}
\affiliation{Center for Computational Quantum Physics (CCQ), Flatiron Institute, New York, NY, USA}

\author{Angel Rubio}
\affiliation{Max Planck Institute for the Structure and Dynamics of Matter, Hamburg, Germany, EU}
\affiliation{Center for Computational Quantum Physics (CCQ), Flatiron Institute, New York, NY, USA}

\author{Iva B\v rezinov\'a}
\email{iva.brezinova@tuwien.ac.at}
\affiliation{Institute for Theoretical Physics, Vienna University of Technology,
    Wiedner Hauptstra\ss e 8-10/136, 1040 Vienna, Austria, EU}
\affiliation{Center for Computational Quantum Physics (CCQ), Flatiron Institute, New York, NY, USA}
\date{\today}

\begin{abstract}
We explore the non-equilibrium dynamics of a one-dimensional Fermi-Hubbard system as a sensitive testbed for the capabilities of the time-dependent two-particle reduced density matrix (TD2RDM) theory to accurately describe time-dependent correlated systems. We follow the time evolution of the out-of-equilibrium finite-size Fermi-Hubbard model initialized by a quench over extended periods of time. By comparison with exact calculations for small systems and with matrix product state (MPS) calculations for larger systems but limited to short times, we demonstrate that the TD2RDM theory can accurately account for the non-equilibrium dynamics in the regime from weak to moderately strong inter-particle correlations. We find that the quality of the approximate reconstruction of the three-particle cumulant (or correlation) required for the closure of the equations of motion for the reduced density matrix is key to the accuracy of the numerical TD2RDM results. We identify the size of the dynamically induced three-particle correlations and the amplitude of cross correlations between the two- and three-particle cumulants as critical parameters that control the accuracy of the TD2RDM theory when current state-of-the art reconstruction functionals are employed.
\end{abstract}
\maketitle

\section{Introduction}\label{sec:intro}
Accurately describing the correlated out-of-equilibrium dynamics of interacting many-particle systems has remained a great challenge to date. Frequent realizations of such out-of-equilibrium dynamics involve either quenches and relaxation of initially prepared excited states of systems governed by a time-independent Hamiltonian, or systems driven by an explicitly time-dependent Hamiltonian. Such systems are at the forefront of current experimental and theoretical studies (see e.g.~ \cite{pohl_2000, cazlilla_2002, caillat_2005, burnus_2005, otobe_2008, eckstein_thermalization_2009, driscoll_2011, hochstuhl_2012,hochstuhl_2014, wachter_2014,sato_communication_2018, pedersen_symplectic_2019,topp_all-optical_2018,buzzi_higgs-mediated_2021}). Several recent experiments have shown that exotic states of matter can be generated by ultrashort pulses of external fields or energetic ions and that relaxation and decoherence can be strongly influenced by inter-particle correlations  \cite{stojchevska_ultrafast_2014,schlunzen_2016,giannetti_ultrafast_2016,balzer_2018,basov_polariton_2020,kennes_moire_2021,budden_evidence_2021,niggas_2022,bloch_strongly_2022}.\\
A versatile method to reliably describe the non-equilibrium scenarios of correlated many-body systems, in particular in extended systems and for extended periods of time, is still lacking. Direct many-body wavefunction based methods can be applied only to systems with a moderate number of degrees of freedom and pure states as they eventually face the exponential wall of computational effort when increasing the number of particles and the time interval of propagation \cite{zanghellini_2003, hochstuhl_2014,cazlilla_2002,haegeman_time-dependent_2011}. Application of the time-dependent density matrix renormalization group (DMRG) theory \cite{cazlilla_2002,daley_time-dependent_2004} has been shown to yield numerically accurate results, currently, however, limited to one-dimension (1D) systems and short time scales (see e.g.~\cite{kollath_2005,schlunzen_2017, joost_dynamically_2022}). Similarly, the closely related time-dependent matrix product state (MPS) method \cite{haegeman_time-dependent_2011,haegeman_unifying_2016} invoking the time-dependent variational principle, is also limited to small propagation times for mesoscopic system sizes of a few tens of particles (see e.g.~\cite{kloss_time-dependent_2018}) with the increasing bond dimension as a function of time as the major bottleneck (see e.g.~\cite{yang_2020}).\\
The complex multi-dimensional information encoded in the quantum many-body wavefunction is, however,  often not needed for the extraction of many physical observables. Therefore, an appealing alternative are time-dependent quantum many-body methods that attempt to bypass the use of the many-body wavefunction altogether. Upon successively tracing out more and more degrees of freedom, information and complexity is lost but, in turn, the reduced system is rendered increasingly tractable.\\
A well-known limit of this reduction is the time-dependent particle density $n(\bm{r},t)$. The corresponding many-body theory, the time-dependent density functional theory (TDDFT) \cite{runge_1984,ullrich_2012} with the Kohn-Sham ansatz features a linear scaling with particle number and remains to date the only time-dependent quantum many-body theory applicable to large extended systems with weak to intermediate correlations. Its major drawback, however, is the fundamental lack of knowledge of the exact exchange-correlation (XC) functional. The pathway towards systematic improvements beyond the currently frequently used approximate adiabatic XC functionals is still a widely open question and the applicability of TDDFT to correlated systems is limited. Alternatively, the so-called time-dependent current-density functional theory has been proposed for which, up to now, however only few approximations for the
exchange-correlation vector potential have become available \cite{vignale_current-dependent_1996, dagosta_relaxation_2006, furness_current_2015}.\\
Going up one step of the ladder of reduction the one-particle reduced density matrix (1RDM) $D_1(\bm{r}_1,\bm{r}'_1,t)$ allows one to avoid some of the problems of TDDFT \cite{pernal_2007, giesbertz_charge_2008, giesbertz_response_2010} while facing others. The equation of motion for the 1RDM corresponds to the first equation within the Bogoliubov-Born-Green-Kirkwood-Yvon (BBGKY) hierarchy \cite{huang_2008,bonitz_2015} and thus couples the 1RDM to the two-particle reduced density matrix (2RDM) $D_{12}(\bm{r}_1,\bm{r}_2,\bm{r}'_1,\bm{r}'_2,t)$. Closing the equations of motion requires representing the 2RDM as a functional of the 1RDM which is challenging in the presence of medium to strong correlations and time-dependent settings. \\
An alternative route to an accurate description of non-equilibrium correlated quantum many-body systems involves non-equilibrium Green's function (NEGF) methods, going back to the pioneering work of Keldysh \cite{keldysh_1965}. They have been applied to a wide range of physical systems (see e.g.~\cite{stefanucci_2013, schlunzen_ultrafast_2020} and references therein) but are impeded by a non-linear time scaling, which has only recently been overcome \cite{schlunzen_achieving_2020, joost_dynamically_2022,karlsson_2021,pavlyukh_2022}. Moreover, they exhibit a similar hierarchical coupling between different orders of Green's functions, which is subject to closure approximations as in the case of reduced density matrices (see e.g.~\cite{stefanucci_2013}).\\
The importance of two-particle correlations as imprinted by the pair-wise interaction potentials in most physical systems calls for the use of the 2RDM itself as the fundamental object for representing the many-body system. When only one- and two-body operators are present in the Hamiltonian, the total energy of the system can be \textit{exactly} expressed in terms of the 2RDM. The fact that the energy is an exactly known functional of the 2RDM has been meanwhile exploited in numerous calculations of groundstate energies in quantum many-body systems \cite{mazziotti_realization_2004,mazziotti_variational_2006, hammond_variational_2006, deprince_parametric_2007, nakata_variational_2008}. \\
In this paper we investigate the time-dependent 2RDM. The equation of motion for propagating the 2RDM of an excited system, the second equation of motion the BBGKY hierarchy, requires, the knowledge of the three-particle reduced density matrix (3RDM). Many important works have been devoted in the past to develop reconstruction functionals of the 3RDM in terms of the 2RDM for the quantum many-body ground state problem \cite{colmenero_approximating_1993, yasuda_direct_1997, mazziotti_pursuit_1999,mazziotti_complete_2000, deprince_cumulant_2007, tohyama_truncation_2017, tohyama_truncation_2019}. Incorporating such reconstruction functionals into the time-depending setting within the time-dependent 2RDM method (TD2RDM), we have recently succeeded in calculating the dynamics of multi-electron atoms driven by strong laser fields \cite{lackner_propagating_2015,lackner_high-harmonic_2017}. Motivated by the stability and remarkable accuracy of this method, it is the aim of the present paper to explore the application of the TD2RDM theory to extended systems, and to systems featuring stronger correlations than typically present in multi-electron atoms.\\
A paradigmatic model system for this endeavor is the Fermi-Hubbard model due to its structural simplicity and the one-parameter tunability from weakly to strongly correlated dynamics. Moreover, this model system can nowadays be realized and accurately probed with ultracold atoms in optical lattices even with single-site resolution (see e.g.~\cite{haller_single-atom_2015,greif_site-resolved_2016, parsons_site-resolved_2015,cheuk_quantum-gas_2015,cheuk_observation_2016, chiu_quantum_2018,eisert_quantum_2015} and references therein) and is, of course, of conceptual relevance for the study of correlated quantum matter in real solids. Several state-of-the art methods have been tested by application to the Fermi-Hubbard model. They include the NEGF methods \cite{hermanns_2014,schlunzen_2016, schlunzen_2017}, as well as approaches based on Green's functions exploiting the mapping between the Fermi-Hubbard model and an impurity model where the impurity is treated in a fully correlated fashion and is coupled to an external uncorrelated bath. These methods, such as time-dependent dynamical mean-field theory \cite{eckstein_thermalization_2009} or the explicit sum of a high-order perturbation series in the interaction on the Keldysh contour using quantum Monte-Carlo methods \cite{bertrand_quantum_2019, bertrand_reconstructing_2019, nunez_fernandez_learning_2022} have the advantage that extended systems can be treated through the coupling of the impurity to an extended bath. However, correlations between distant sites are not well represented.\\
As a prototypical example, we apply the TD2RDM theory to the dynamics of the Fermi-Hubbard model at half filling initialized by a quench, i.e.~by suddenly switching off a confining potential that prepares the initial out-of-equilibrium state (Fig.~\ref{fig:setup}). In order to test and to benchmark the TD2RDM we consider in the present work one-dimensional systems with a relatively small number of sites. For these systems a detailed assessment of the accuracy by comparison with numerically exact or highly accurate solutions is still possible allowing us to perform large and systematic parameter scans over many different interaction strengths and excitation energies. We generate (nearly) exact solutions by direct propagation of the Schr\"odinger equation or using highly accurate matrix product state calculations (MPS) within the time-dependent variational principle \cite{haegeman_time-dependent_2011, haegeman_unifying_2016, kloss_time-dependent_2018}. We follow the dynamics over relatively long times ($\geq 50$ in units of the inverse hopping amplitude) and study the exact build-up of dynamical correlations which can give valuable hints for the applicability of the TD2RDM method as well as for the improvements of reconstruction functionals. We emphasize that our present restriction to 1D systems of moderate size is due to the difficulty of obtaining exact or highly accurate results for comparison, rather than due to the limitations of the TD2RDM theory itself. The latter can be easily extended to larger systems and higher dimensions without encountering major complications. We also compare with time-dependent Hartree-Fock (TDHF) predictions to access the influence of two-particle correlations neglected by mean-field theories. We analyze the accuracy of the TD2RDM theory as a function of the strength of the inter-particle interaction as well as the degree of initial excitation. Our focus is on detailed probes of the accuracy of the time-dependent three-particle correlations resulting from different state-of-the art reconstruction functionals. \\
The structure of the paper is as follows: In Sec.~\ref{sec:sys} we briefly present the model system under investigation, the one-dimensional Fermi-Hubbard model at half filling. The key ingredients of the TD2RDM theory are reviewed in Sec.~\ref{sec:theory}. We numerically analyze the dynamics of two- and three-particle correlations, the so-called cumulants, which are the key ingredient to reconstruction functionals, for small systems by comparison with exact calculations in Sec.~\ref{sec:cumu_dyn}. Fully self-consistent TD2RDM simulations for the time evolution of the out-of-equilibrium dynamics as monitored by the one-site occupation number are presented in Sec.~\ref{sec:res_bench}, followed by concluding remarks and an outlook to future improvements in Sec.~\ref{sec:concl}. As units we use $\hbar=m=e=1$ unless otherwise stated.
\section{Out-of-equilibrium Fermi-Hubbard model}\label{sec:sys}
\begin{figure}[t]
	\includegraphics[width=\columnwidth]{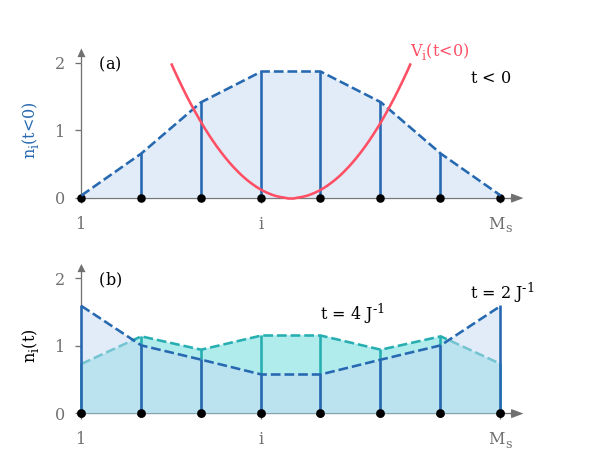}
	\caption{Fermi-Hubbard model with $M_s$ sites and external harmonic potential ($t<0$). (a) The one-particle site-occupation number $n_i$ of the initial ground state in the potential representing an excitation of the potential-free Fermi-Hubbard model after the quench. (b) Snapshots of the time evolved $n_i(t)$ at $t=2  J^{-1}$ and $t = 4J{-1}$. The parameters used are $V= J$ and $U=J$.}
	\label{fig:setup}
\end{figure}

We consider a 1D chain with a number of $M_s$ sites (Fig.~\ref{fig:setup}) and impose Dirichlet boundary conditions. 
The Hamiltonian of the Fermi-Hubbard model in the presence of an external potential initializing the quench is given in second quantization by 
\begin{equation}
	H = -J\sum_{\langle i,j\rangle}\sum_{\sigma} a_{i\sigma}^\dagger a_{j\sigma} + U \sum_i n_i^{\uparrow}n_i^{\downarrow} + \sum_{i,\sigma} V_i(t) a^{\dagger}_{i\sigma}a_{i\sigma},
	\label{eq:ham_sec}
\end{equation}
where $\langle i,j\rangle$ denotes nearest-neighbor hopping, $J$ the hopping amplitude, $n_i^{\uparrow (\downarrow)} = a_{i\uparrow(\downarrow)}^\dagger a_{i\uparrow(\downarrow)}$, the occupation number operators for particles with spin up (down) at site $i$, $U$ the strength of on-site interaction controlling the correlation energy in the system. $V_i(t)$ is the explicitly time dependent potential chosen to be harmonic in the present case,
\begin{equation}
	V_i(t) = \theta(-t)\frac{V^2}{2}\left(i-\frac{M_s+1}{2}\right)^2 ,
	\label{eq:harm_pot}
\end{equation}
which determines the initial excited state (the ground state of $H$ in the potential for $t<0$), and induces the dynamics by a sudden potential quench at $t=0$. We consider in the following the spin-symmetric Fermi-Hubbard system at half filling, i.e.~particle number $N=M_s$ and the number of spin up particles equals to the number of spin down particles (total spin-singlet case). \\
Fig.~\ref{fig:setup} illustrates the quench-induced dynamics on the level of the one-particle site-occupation number $n_i$ corresponding to the diagonal elements of the one-particle reduced density matrix $D_1$. The ground state of the interacting many-body system in the potential [Fig.~\ref{fig:setup} (a)] represents an excited state of the field-free Fermi-Hubbard system and, thus, an out-of-equilibrium state that evolves in time after the quench [Fig.~\ref{fig:setup} (b)]. It would eventually relax,upon coarse graining, to a new equilibrium state. We explore in the following within the framework of TD2RDM theory the importance of inter-particle correlations induced by $U$ (Eq.~\ref{eq:ham_sec}) in both the stationary initial state as well the time-dependent correlations induced by the sudden quench.
\section{Outline of TD2RDM theory}\label{sec:theory}
\subsection{Equation of motion}\label{subsec:eom}
The central object of our method is the 2RDM which is obtained from the exact pure $N$-body wavefunction $|\Psi(t)\rangle$ by tracing out all but two particles. We denote the 2RDM in a basis-independent notation as $D_{12}$ and it follows from  $|\Psi(t)\rangle$ as 
\begin{equation}
	D_{12} (t) = N(N-1) {\rm Tr}_{3\hdots N} |\Psi(t)\rangle \langle \Psi(t)|,
	\label{eq:D12_tr}
\end{equation}
with $N$ the number of particles, $N(N-1)$ the normalization related to particles pairs, and ${\rm Tr}_{3\hdots N}$ indicating the tracing out of all particles except for the two particles $1$ and $2$ of interest. More generally, the pRDM is obtained from 
\begin{equation}
	D_{1\hdots p} (t) = \frac{N!}{(N-p)!} {\rm Tr}_{p+1\hdots N} |\Psi(t)\rangle \langle \Psi(t)|,
	\label{eq:DpN_tr}
\end{equation}
with normalization factor $N!/(N-p)!$.\\
The equation of motion of the 2RDM corresponds to the second equation within the BBGKY hierarchy and reads 
\begin{align}
	i\partial_t D_{12}(t) = &[h_1+h_2+W_{12},D_{12}] \nonumber \\
	&+ {\rm Tr}_3[W_{13}+W_{23},D_{123}]
	\label{eq:eom}
\end{align}
where the square brackets denote commutators. The Hamiltonian governing Eq.~\ref{eq:eom}  is given (in first quantization) by 
\begin{equation}
	H = \sum_{n=1}^N h_n + \sum_{n<m}^N W_{nm},
	\label{eq:hamilton}
\end{equation}
where $h_n$ is the single-particle Hamilton operator, and $W_{nm}$ the two-particle interaction operator. In a basis of spin orbitals $\{|\psi_{i\sigma}\rangle \}_{i=1}^{M_s}$ with $\sigma=\uparrow$ or $\sigma=\downarrow$ localized at a single site $i$ (given e.g.~by s-wave orbitals localized at atomic sites in solids or potential minima in optical latices of ultracold atoms) the terms in Eq.~\ref{eq:hamilton} yield the explicit matrix representation for the nearest neighbor hopping as
\begin{align}
	h_{j\sigma}^{i\sigma'} &= \langle \psi_{i\sigma'}| h_1 | \psi_{j\sigma}\rangle
	\nonumber \\
	&=-J\delta_{j}^{i+1}\delta_{\sigma}^{\sigma'} - J\delta_{j}^{i-1}\delta_{\sigma}^{\sigma'},
	\label{eq:h_1}
\end{align}
and the on-site interaction of particles with different spins as
\begin{align}
	W_{j_1\sigma_1j_2\sigma_2}^{i_1\sigma_1'i_2\sigma_2'} &=
	\langle \psi_{i_1\sigma'} \psi_{i_2\sigma'}| W_{12}| \psi_{j_1\sigma} \psi_{j_2\sigma}\rangle
	\nonumber \\
	&= U
	\delta_{j_1}^{i_1}\delta_{j_2}^{i_2}\delta_{j_1,j_2}
	\delta_{\sigma_1}^{\sigma_1'}\delta_{\sigma_2}^{\sigma_2'}(1-\delta_{\sigma_1,\sigma_2}).
	\label{eq:h_int}
\end{align}
For any initial state (pure or mixed) described by $D_{12}(t=0)$, Eq.~\ref{eq:eom} allows to propagate the 2RDM without any knowledge of the many-body wavefunction $|\Psi(t)\rangle$. However, since all equations of the BBGKY hierarchy couple to the density matrix of the next higher order, propagation of the 2RDM requires closure, i.e.~a sufficiently accurate representation for the 3RDM in terms of the 2RDM. Closure of the equations of motion by reconstruction (denoted by the superscript R in the following) of the 3RDM by the 2RDM, i.e.
\begin{equation}
	D_{123} \approx D_{123}^\text{R}[D_{12}]
	\label{eq:D123_rec}
\end{equation}
poses thus a major challenge for the implementation of the TD2RDM theory as a useful and accurate computational tool. In the spirit of a quantum Boltzmann transport equation \cite{huang_2008}, we call the term in Eq.~\ref{eq:eom} containing the $D_{123}$ the collision operator (or ``collision integral") $C$, 
\begin{equation}
	C[D_{123}] = {\rm Tr}_3[W_{13}+W_{23}, D_{123}].
\end{equation}
While for the collision operator an approximation to the reconstruction of the 3RDM is required, the time-dependent 2RDM, $D_{12}(t)$, fully includes all two-particle interactions and correlations without any additional approximation. The solutions of the equations of motion of the 2RDM (Eq.~\ref{eq:eom}) feature an important exact relation to Green's functions which opens the door to employ well established diagramatic methods also within the TD2RDM theory. The pRDMs can be identified with the equal-time limits of the p-particle Green's functions $G^{<}_{1\hdots p}(t_1,\hdots, t_p,t'_1,\hdots, t'_p)$.
For the 1RDM and 2RDM, e.g., we get (see e.g.~\cite{joost_dynamically_2022, stefanucci_2013})
\begin{align}
	D_1(t) &= -\lim_{\delta\rightarrow 0}iG^{<}_1(t,t+\delta) \\
	D_{12}(t) &= \lim_{\delta\rightarrow 0}i^2G^{<}_{12}(t,t,t+\delta,t+\delta).
\end{align}
In a given single particle basis $D_{12}(t)$ is represented by the matrix
\begin{equation}
	D_{j_1\sigma_1j_2\sigma_2}^{i_1\sigma'_1i_2\sigma'_2} = \langle \Psi(t)|a_{i_1\sigma'_1}^\dagger a_{i_2\sigma'_2}^\dagger a_{j_2\sigma_2}a_{j_1\sigma_1}|\Psi(t)\rangle.
	\label{eq:D12_orb_bas_sig}
\end{equation}
Because of the dependence of $C$ on $D_{123}$, the equation of motion of the 2RDM represented in a single-particle basis of dimension $M$ scales as $M^7$ for a general pair-interaction $W$. 
In the present spin-symmetric realization of the Fermi-Hubbard model with equal number of spin-up and spin-down particles, the complexity of the problem can be considerably reduced. The calculation of Eq.~\ref{eq:eom} can be reduced to that of the spin block $D_{j_1\uparrow j_2\downarrow}^{i_1\uparrow i_2\downarrow}$ which contains all the information on the entire $D_{12}(t)$. All other spin blocks can be obtained from this particular block either through trivial exchange or spin-flip symmetries, or through the following relation
\begin{equation}
	D_{j_1\uparrow j_2\uparrow}^{i_1\uparrow i_2\uparrow} =
	D_{j_1\uparrow j_2\downarrow}^{i_1\uparrow i_2\downarrow} 
	- 	D_{j_2\uparrow j_1\downarrow}^{i_1\uparrow i_2\downarrow}  .
\end{equation}
Correspondingly, only the 3RDM block $D_{j_1\uparrow j_2\uparrow j_3\downarrow}^{i_1\uparrow i_2\uparrow i_3\downarrow}$ needs to be constructed instead of the entire 3RDM. The equation of motion for $D_{j_1\uparrow j_2\downarrow}^{i_1\uparrow i_2\downarrow}$ is given in Appendix \ref{app:eom}. Due to the simple on-site interaction within the Fermi-Hubbard model (Eq.~\ref{eq:h_int}), the equation of motion for the 2RDM scales as $M_s^4$. For simplicity of notation, we drop the explicit spin labeling ($\uparrow, \downarrow$) unless specifically needed keeping in mind that only the spin blocks identified above need to be calculated. 
\subsection{Cumulant expansion}\label{subsec:cumu}
\begin{figure}[t]
	\includegraphics[width=\columnwidth]{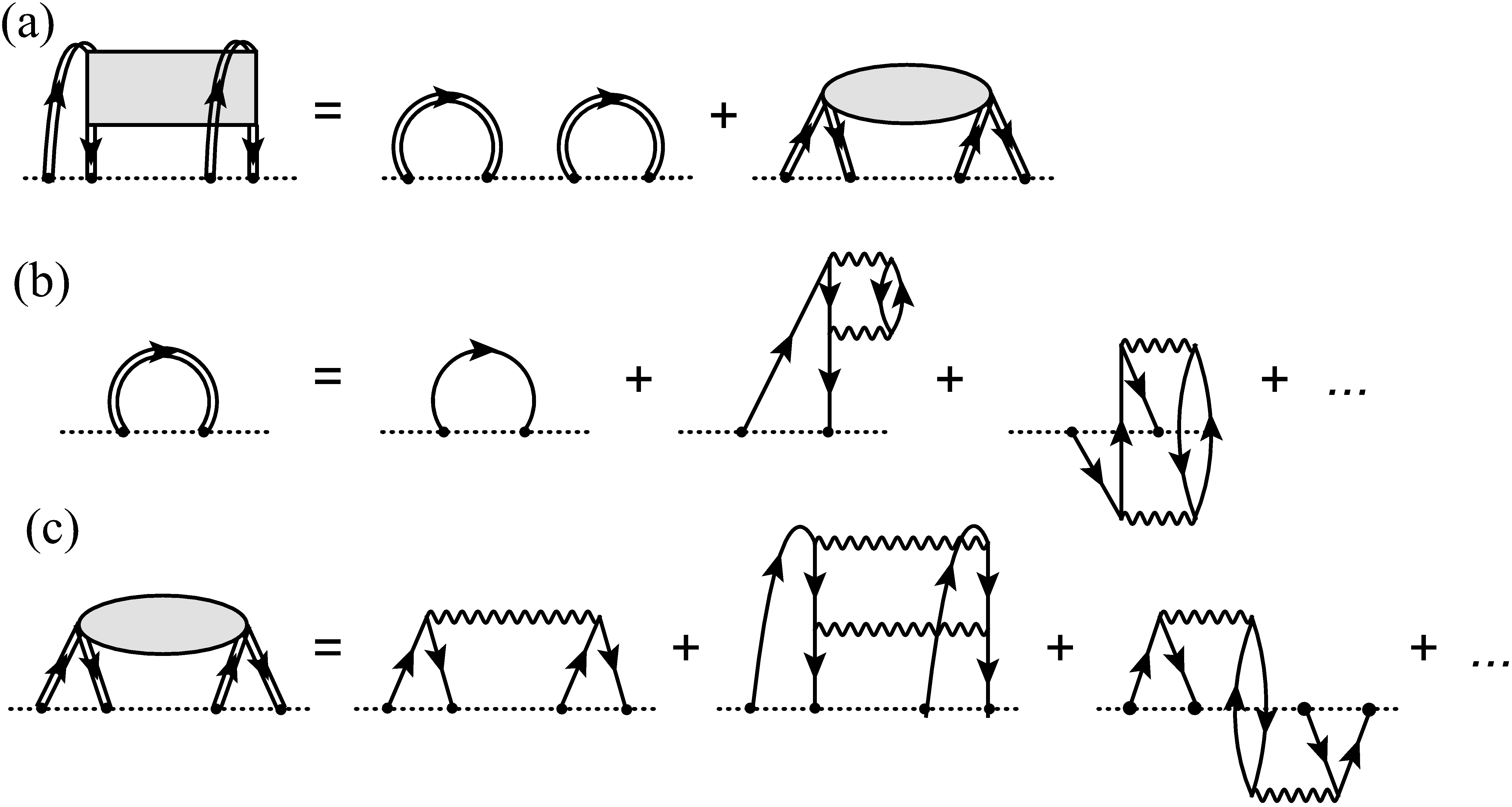}
	\caption{(a) Diagrammatic representation of the 2RDM. The first term corresponds to $\hat A D_1D_2$ (the anti-symmetrized contribution is not explicitly shown for brevity). The second term represents the cumulant $\Delta_{12}$ which contains all connected diagrams between two single-particle propagators. Note that no perturbative expansion in the inter-particle interaction $W_{12}$ is involved in (a). (b) Diagrammatic perturbative expansion of $D_1$, i.e.~the equal-time limit of the full single-particle propagator. Two second order diagrams in $W_{12}$ are shown as illustrative examples. The double line represents the full $D_1$, while each single line in (b) and (c) stands for a Hartree-Fock propagator. (c) Diagramatic perturbative expansion of the cumulant $\Delta_{12}$ with three prototypical diagrams to first and second order in $W_{12}$.}
	\label{fig:diags_2rdm}
\end{figure}

\begin{figure}[t]
	\includegraphics[width=\columnwidth]{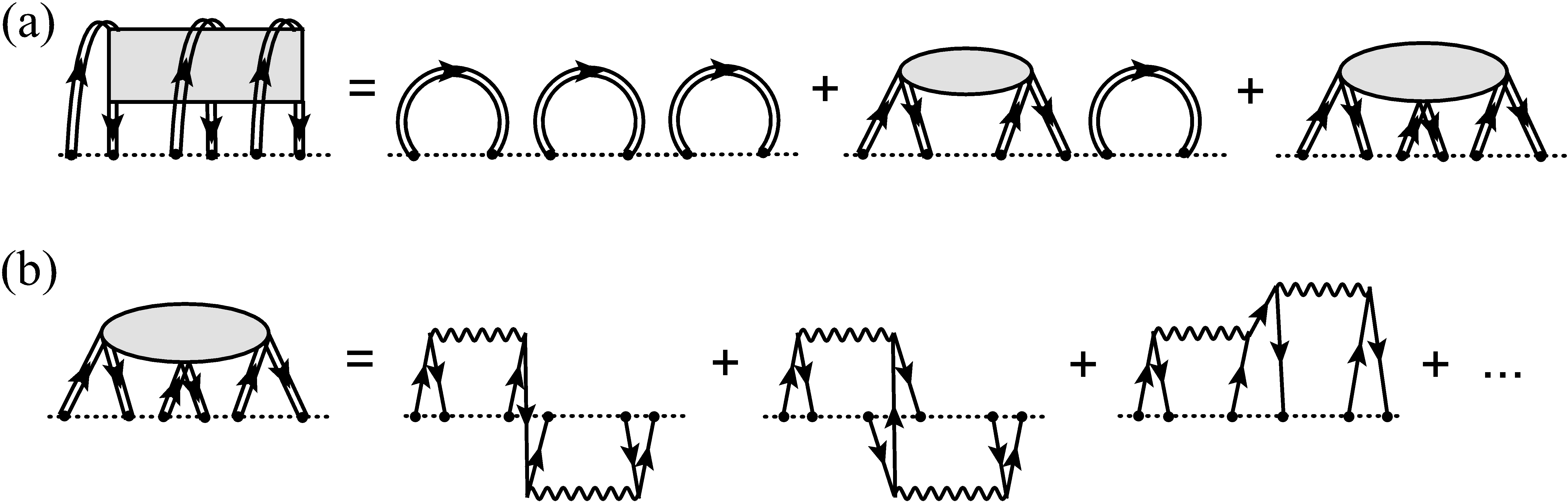}
	\caption{(a) Diagrammatic representation of the 3RDM. The first term corresponds to $\hat A D_1D_2D_3$, the second corresponds to $\hat A \Delta_{12}D_3$, and the last to the three-particle cumulant $\Delta_{123}$. The anti-symmetrized contributions of each term are not shown for brevity. (b) Diagrammatic representation of the perturbative expansion of the three-particle cumulant, with several prototypical second/order connected diagrams ($\propto W_{12}^2$) shown.}
	\label{fig:diags_3rdm}
\end{figure}

The pRDM describes, in general, the correlated dynamics of a $p$-tuple of particles embedded in a larger system, in particular in the pure state $|\Psi(t)\rangle$ of an $N$-particle system. In the absence of inter-particle interactions, the pRDM reduces to the independent-particle limit where only Pauli exchange correlations via anti-symmetrization are present. Accordingly, the pRDM can be expanded in term of correlators, in this context conventionally referred to as cumulants \cite{kutzelnigg_cumulant_1999}, of increasing order in the number of particles within the tuple to be correlated with each other.\\
For $D_{12}$ the cumulant expansion reads 
\begin{equation}
	D_{12} = \hat A D_1D_2 + \Delta_{12}
	\label{eq:2rdm_cum}
\end{equation}
with the two-particle cumulant (or correlator) $\Delta_{12}$ and $\hat A$ the anti-symmetrization operator acting on the two one-particle density matrices $D_1$ and $D_2$. In the single-particle site representation
\begin{equation}
	\hat A D_{j_1}^{i_1}D_{j_2}^{i_2} = D_{j_1}^{i_1}D_{j_2}^{i_2}-D_{j_2}^{i_1}D_{j_1}^{i_2}.
\end{equation}
and Eq.~\ref{eq:2rdm_cum} reads
\begin{equation}
	D_{j_1j_2}^{i_1i_2} = \hat A D_{j_1}^{i_1}D_{j_2}^{i_2}  + \Delta_{j_1j_2}^{i_1i_2}.
	\label{eq:2rdm_cumu_dec}
\end{equation}
The cumulant expansion of $D_{12}$ (Eq.~\ref{eq:2rdm_cum}) can be diagrammatically visualized (Fig.~\ref{fig:diags_2rdm}). 
The key feature to be noted is that the cumulant expansion [Fig.~\ref{fig:diags_2rdm} (a)] does not invoke any ingredients from perturbation theory. The double lines represent the equal-time limit of the full one-particle propagator. The cumulant represents the sum over all connected diagrams between two one-particle propagators. For illustrative purposes and to connect to other theories we also indicate in Fig.~\ref{fig:diags_2rdm} (b) and (c) the corresponding perturbative diagrammatic expansion of the constituents of Fig.~\ref{fig:diags_2rdm} (a), the one-particle propagator [Fig.~\ref{fig:diags_2rdm} (b)] and the two-particle cumulant [Fig.~\ref{fig:diags_2rdm} (c)]. We emphasize that within the TD2RDM theory the full 1RDM as well as the full 2RDM are included such that the use of the perturbation series [Fig.~\ref{fig:diags_2rdm} (b), (c)] can be avoided. However, these diagrammatic interrelations provide a helpful guidance for developing reconstruction functionals on the three-particle level.\\
The cumulant expansion of the 3RDM follows as 
\begin{equation}
	D_{123} = \hat A D_1 D_2 D_3 + \hat A \Delta_{12}D_3 + \Delta_{123},
	\label{eq:3rdm_cumu}
\end{equation}
diagrammatically visualized in Fig.~\ref{fig:diags_3rdm} (a). 
The first term in Fig.~\ref{fig:diags_3rdm} (a) represents three uncorrelated particles, the second the contribution of two-particle correlations in the presence of a third uncorrelated particle, and the last the true three-particle correlation or three-particle cumulant $\Delta_{123}$ containing all connected three-particle diagrams. For illustrative purposes we show also in Fig.~\ref{fig:diags_3rdm} (b) the first few low-order diagrams of a perturbative expansion of $\Delta_{123}$ in terms of Hartree-Fock propagators and pair interactions. We note again that the present TD2RDM theory does not make direct use of perturbation theory but we invoke the structure of these diagrams in the following to motivate the approximations of $\Delta_{123}$ in terms of one-particle propagators and two-particle cumulants.
\subsection{Three-particle cumulant reconstruction}\label{subsec:3cumu_rec}
The challenge to render the TD2RDM theory operational is the closure of the equations of motion (Eq.~\ref{eq:eom}) by developing a reconstruction functional for the three-particle density matrix $D_{123}^\text{R}[D_{12}]$ (Eq.~\ref{eq:D123_rec}). The success of the TD2RDM method in describing the many-body dynamics relies on a sufficiently accurate approximation of this functional as has been shown for multi-electron atoms \cite{lackner_propagating_2015, lackner_high-harmonic_2017}. While for the non-degenerate ground state the existence of such a reconstruction is assured through Rosina's theorem \cite{rosina_book_1968, mazziotti_book_2007}, it is presently unknown, whether such an exact reconstruction also exists in a time dependent setting. As Rosina's theorem is an existence theorem, it does not lend itself to aid in the development of new functionals.\\
The cumulant expansion of $D_{123}$ (Eq.~\ref{eq:3rdm_cumu}) reduces the task of finding a reconstruction functional to that of reconstructing the cumulant $\Delta_{123} = \Delta_{123}^\text{R}[D_{12}]$ as the other terms contributing to $D_{123}$ are already known functionals of $D_{12}$ (and $D_1$). Several approximate functionals $D_{123}^\text{R}[D_{12}]$ or $\Delta_{123}^\text{R}[D_{12}]$ have been recently proposed \cite{colmenero_approximating_1993,yasuda_direct_1997, mazziotti_pursuit_1999, tohyama_truncation_2017,tohyama_truncation_2019}. They provide the starting point of our analysis of the capability of the TD2RDM theory to capture non-equilibrium dynamics in correlated systems. The simplest approximation attributed to Valdemoro (V) and coworkers \cite{colmenero_approximating_1993} amounts to neglecting $\Delta_{123}$ altogether. Accordingly, the reconstruction functional becomes 
\begin{equation}
	D_{123}^V[D_{12}] = \hat A D_1 D_2 D_3 + \hat A \Delta_{12}D_3.
	\label{eq:3rdm_v}
\end{equation}
A similar approximation has been earlier investigated by Wang and Cassing \cite{wang_explicit_1985}. Reconstruction functionals that include contributions from $\Delta_{123}$ and benchmarked in this paper have been derived from different perspectives but all rely on approximating $\Delta_{123}$ to second order in $\Delta_{12}$. Nakatsuji and Yasuda (NY) \cite{yasuda_direct_1997} used diagrammatic techniques to arrive at 
\begin{equation}
	\Delta_{123}^{\rm NY}[\Delta_{12}] = \hat A\Delta_{12}P_2\Delta_{23},
	\label{eq:cumu3_ny}
\end{equation}
where the intermediate single-particle projector $P_{i}$ is given by 
\begin{equation}
	P_i = (2\Gamma_i-I_i-D_i)^{-1},
	\label{eq:p2_1}
\end{equation}
with $I_i$ the identity matrix, $\Gamma_i$ a diagonal matrix in the eigenrepresentation of the 1RDM with eigenvalues $1$ for the lowest $N$ natural orbitals and zero otherwise. $\Gamma_i$ is frequently (in ground state calculations) referred to as the Hartree-Fock reference matrix. We note, however, that in the present context $\Gamma_i$ refers to the natural orbitals of the non-perturbative 1RDM rather than to mean-field states. It has been shown \cite{mazziotti_pursuit_1999} that this projector can be substantially simplified through an expansion in $\Gamma_i-D_i$, the zeroth order of which yields
\begin{equation}
	P_i = (2\Gamma_i-I_i)^{-1}.	
	\label{eq:p2_2}
\end{equation}
In practice, the summation over the index $2$ in Eq.~\ref{eq:cumu3_ny} is performed in the basis of natural orbitals with a matrix as the projector containing $-1$ for unoccupied and $1$ for occupied orbitals. We have checked that in the regime where the NY approximation is applicable (see Sec.~\ref{sec:res_bench} below), both Eq.~\ref{eq:p2_1} and Eq.~\ref{eq:p2_2} yield very similar results. We, therefore, use the much simpler approximation (Eq.~\ref{eq:p2_2}). Thus, the NY reconstruction functional for $D_{123}$ reads
\begin{align}
	D_{123}^{\rm NY}[D_{12}] &= D_{123}^{\rm V}[D_{12}] + \Delta_{123}^{\rm NY}[\Delta_{12}] \nonumber \\
	&= \hat AD_1D_2D_3 +\hat A\Delta_{12}D_3+\Delta_{123}^\text{NY}[\Delta_{12}].
	\label{eq:3rdm_ny}
\end{align}
%
A similar reconstruction functional suggested by Tohyama and Schuck (TS) \cite{tohyama_truncation_2017, tohyama_truncation_2019} has been derived starting from a coupled-cluster ansatz for the wavefunction. Including an empirically found renormalization factor the TS reconstruction functional amounts to 
\begin{equation}
	\Delta_{123}^{\rm TS} = \frac{1}{\mathcal{N}}\Delta_{123}^{\rm NY},
	\label{eq:cumu3_ts}
\end{equation}
with renormalization factor $\mathcal{N}=1+\frac{1}{4}{\rm Tr_{12}}|\Delta_{12}|^2$. The reconstruction functional thus reads
\begin{equation}
	D_{123}^{\rm TS}[D_{12}] = D_{123}^{\rm V}[D_{12}] + \Delta_{123}^{\rm TS}[\Delta_{12}].
	\label{eq:3rdm_ts}
\end{equation}
Mazziotti (M) devised a similar reconstruction of $\Delta_{123}$ along different lines starting from the cumulant decomposition of the 4RDM and assuming ${\rm Tr}_4\Delta_{1234}=0$ with $\Delta_{1234}$ the four-particle cumulant \cite{mazziotti_pursuit_1999, mazziotti_complete_2000}. This leads to an implicit equation for $\Delta_{123}$ which can be explicitly solved in the eigenbasis of the 1RDM. Further details of this reconstruction functional are summarized in Appendix \ref{app:recon}. The corresponding reconstruction functional of the three-particle cumulant is denoted by $\Delta_{123}^\text{M}$ and the reconstructed three-particle density matrix by 
\begin{equation}
	D_{123}^{\rm M}[D_{12}] = D_{123}^{\rm V}[D_{12}] + \Delta_{123}^{\rm M}[\Delta_{12}]. 
	\label{eq:3rdm_m}
\end{equation}
It has been shown that in the perturbative limit the reconstructions (Eqs.~\ref{eq:3rdm_ny}, \ref{eq:3rdm_m}, \ref{eq:3rdm_ts}) agree with each other to second-order in the inter-particle interaction \cite{deprince_cumulant_2007}.\\
None of the reconstruction functionals presented above preserves, however, important symmetries of the equations of motion (Eq.~\ref{eq:eom}), most importantly the contraction consistency (CC). At each instant of time CC requires 
\begin{equation}
	D_{12}(t) = \frac{1}{N-2}\text{Tr}_3D_{123}^\text{R}(t)
\end{equation}
to hold. We have recently shown \cite{lackner_propagating_2015, lackner_high-harmonic_2017} that the lack of CC seriously impedes the stability as well as the accuracy of the solutions of the equation of motion of the TD2RDM. This deficiency, however, can be cured for any reconstruction functional \cite{lackner_propagating_2015, lackner_high-harmonic_2017} by way of unitary decomposition of tensors. The unitary decomposition allows to separate any p-particle matrix $M_{j_1\hdots j_p}^{i_1\hdots i_p}$ into basis-invariant components
\begin{equation}
	M_{12\hdots p} = M_{12\hdots p;\perp} + M_{12\hdots p; \rm K},
	\label{eq:unit_decomp}
\end{equation}
where $M_{12\hdots p; \rm K}$ denotes the kernel $M_{12\hdots p; \rm K}$ under contractions, i.e.
\begin{equation}
	\text{Tr}_pM_{12\hdots p; \rm K} = 0,
\end{equation}
while $M_{12\hdots p;\perp}$ denotes the component orthogonal to the kernel. $M_{12\hdots p;\perp}$ carries all the important information encoded in $M_{12\hdots p}$ that survives in the lower dimensional space upon contraction. In turn, $M_{12\hdots p;\perp}$ can be reconstructed from the information available in the contracted space. Eq.~\ref{eq:unit_decomp} applied to the 3RDM yields 
\begin{equation}
	D_{123} =  D_{123; \rm K} + D_{123;\perp}[D_{12}],
\end{equation}
with the important consequence that the orthogonal component of $D_{123}$ as well as of $\Delta_{123}$ become now exactly known functionals of the 2RDM.  This exact functional for three-particle hermitian matrices has been first given in \cite{lackner_propagating_2015} (see also \cite{lackner_phd_2017} and \cite{joost_dynamically_2022} for a more detailed description). With this decomposition we can now reconstruct parts of the missing components for the above reconstruction functionals through 
\begin{equation}
	D_{123}^{\rm R+CC}[D_{12}] = D_{123; \rm K}^{\rm R}[D_{12}] +  D_{123;\perp}[D_{12}] 
	\label{eq:rec+CC_1}
\end{equation}
or equivalently
\begin{equation}
	D_{123}^{\rm R+CC}[D_{12}] = D_{123}^{\rm R}[D_{12}] + D_{123;\perp}[D_{12}^{\rm d}],
	\label{eq:rec+CC_2}
\end{equation}
where the defective part of the 2RDM, $D_{12}^{\rm d}$, corresponds to the contraction error in the two-particle space
\begin{equation}
	D_{12}^{\rm d} = D_{12} - \frac{1}{N-2}{\rm Tr}_3D_{123}^{\rm R}.
\end{equation}
By construction, $D_{123}^\text{R+CC}$ is now contraction consistent, i.e.
\begin{equation}
	D_{12} = \frac{1}{N-2}\text{Tr}_3 D_{123}^\text{R+CC}[D_{12}].
\end{equation}
Equally importantly, the CC correction to $D_{123}$, $D_{123;\perp}[D_{12}^{\rm d}]$, provides a correction to the approximate three-particle cumulant
\begin{equation}
	 \Delta_{123;\perp}[D_{12}] = D_{123;\perp}[D_{12}^\text{d}].
	\label{eq:cumu_corr}
\end{equation}
In Eq.~\ref{eq:cumu_corr} we have used the fact that the first two terms of the cumulant expansion (Eq.~\ref{eq:3rdm_cumu}) are already exact functionals of $D_{12}$. One remarkable consequence of restoring parts of $\Delta_{123}$ by the CC correction is that even the Valdemoro approximation whose bare version (Eq.~\ref{eq:3rdm_v}) neglects $\Delta_{123}$ entirely contains now in its contraction consistent (V+CC) version a three-particle correlation contribution $\Delta_{123;\perp}$. The residual error for the reconstruction functionals considered can thus be traced to the kernel of the three-particle cumulant $\Delta_{123;K}$, either completely missing as in the V+CC approximation or only incompletely reconstructed by the NY+CC, TS+CC, or M+CC approximation. In the following, we refer to functionals without the CC correction as the bare functionals.
\section{Probing the dynamics of the cumulants}\label{sec:cumu_dyn}
\begin{figure}[t]
	\includegraphics[width=\columnwidth]{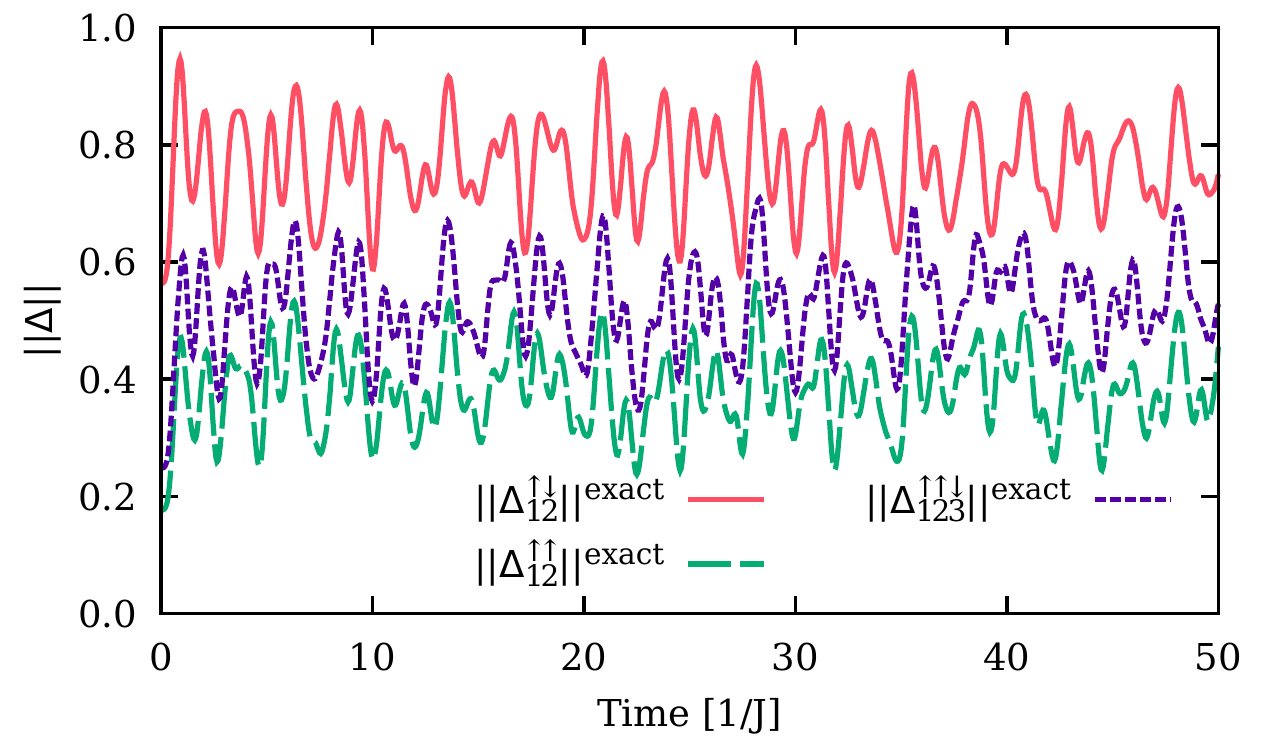}
	\caption{Time evolution of the exact cumulants (Frobenius norm) for the six-site Fermi-Hubbard model at half filling after the potential quench with $V= 1 J$ and $U=3.1 J$. 
	}
	\label{fig:cumu_dyn_examp}
\end{figure}

\begin{figure}[t]
	\includegraphics[width=\columnwidth]{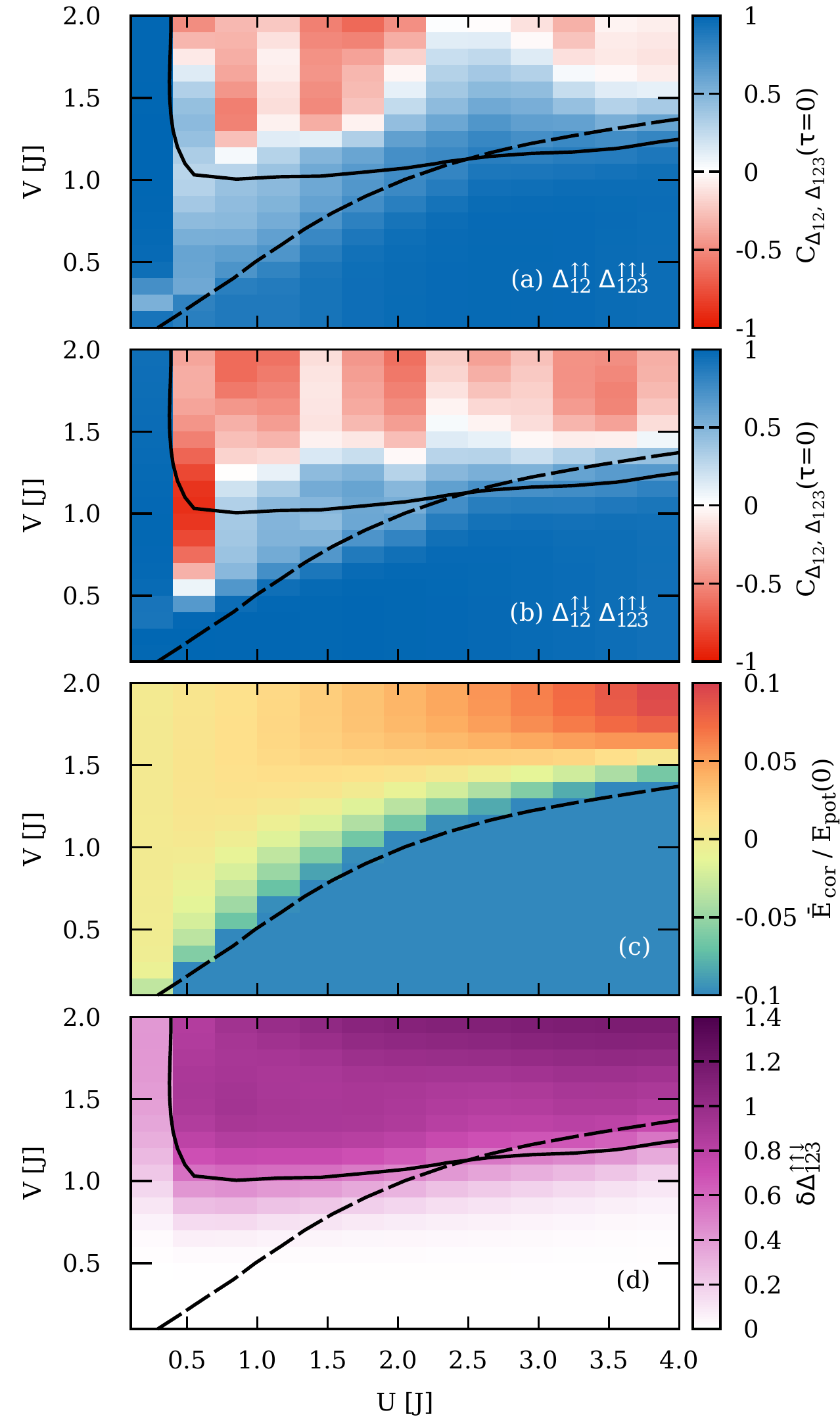}
	\caption{Six-site Fermi-Hubbard model at half filling ($M_s=6$) with interaction $U$ starting from a ground state at varying potential of strength $V$. Equal-time limit of the normalized cross correlation function between the three-particle cumulant $\Delta_{123}^{\uparrow\uparrow\downarrow}$ and the two-particle cumulants (a) $\Delta_{12}^{\uparrow\uparrow}$ and (b) $\Delta_{12}^{\uparrow\downarrow}$ in the $U$-$V$ plane.  In (c) we depict the mean (i.e.~time averaged) correlation energy $\bar E_\text{cor}$ relative to the initial excitation energy $E_\text{pot}(0)$. The colorbar is cut below and above $\pm0.1$ (i.e.~larger and smaller values than shown on the color bar are present.) The black dashed line indicates the $-0.1$ contour line. In (d) we show the dynamical build-up of three-particle correlations relative to the initial correlations at $t=0$, $\delta \Delta_{123}^{\uparrow\uparrow\downarrow}$ (Eq.~\ref{eq:cumu3_build}). The solid black line denotes the $\delta \Delta_{123}^{\uparrow\uparrow\downarrow}=0.65$ contour.}
	\label{fig:pearson_m6}
\end{figure}

\begin{figure}[t]
	\includegraphics[width=\columnwidth]{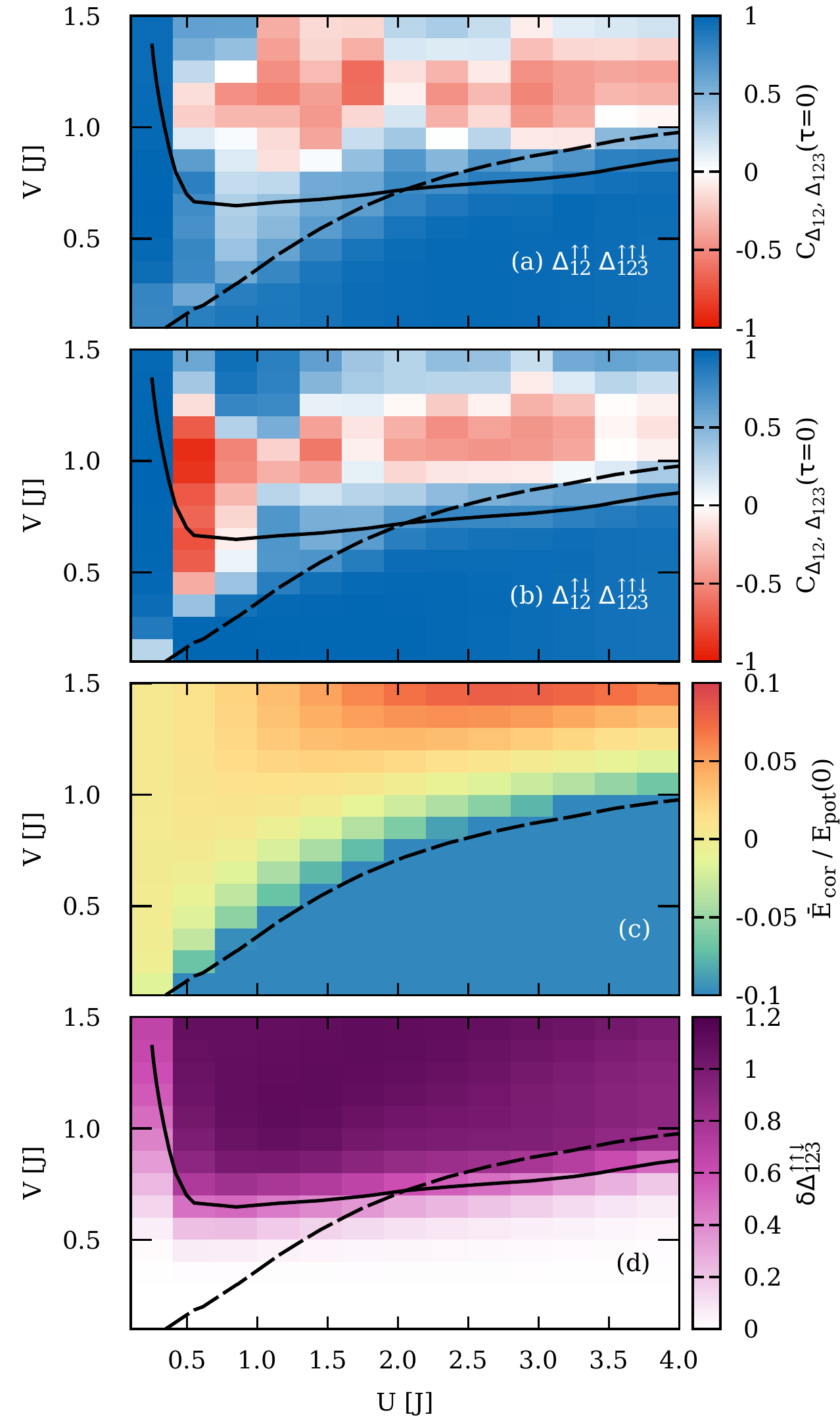}
	\caption{As in Fig.~\ref{fig:pearson_m6}, but for the Fermi-Hubbard model with $M_s=8$ sites.
	}
	\label{fig:pearson_m8}
\end{figure}

The proposed approximate reconstruction functionals for $D_{123}(t)$ or, more specifically, for the three-particle cumulant $\Delta_{123}(t)$ (Eqs.~\ref{eq:cumu3_ny}, ~\ref{eq:cumu3_ts}, ~\ref{eq:3rdm_m}) are at most quadratic functionals in $\Delta_{12}(t)$ and local in time. Higher-order terms in $\Delta_{12}$ as well as any memory effects are neglected from the outset. As this simple analytic structure of the approximate reconstruction functionals implies strong temporal correlations between $\Delta_{12}(t)$ and $\Delta_{123}(t)$, it is instructive to probe for the temporal correlations between the time evolution of $\Delta_{123}(t)$ and $\Delta_{12}(t)$ in the non-equilibrium few-site Fermi-Hubbard model. Only when such time-correlated dynamics is present within the exact solution, the reconstruction by the time-local reconstruction functionals used here can be expected to be accurate.\\
For the Fermi-Hubbard model we explore the coupling between $\Delta_{12}(t)$ and $\Delta_{123}(t)$ by following the quench dynamics for varying strength of interparticle correlations (Hubbard parameter $U$) and strength of the initial out-of-equilibrium excitation (controlled by the confining potential parameter $V$). We extract from the exact propagation the time evolution of $\Delta_{12}(t)$ and $\Delta_{123}(t)$ for the quench dynamics without invoking any reconstruction functional, starting from the exact ground state in the potential well of strength $V$ [Fig.~\ref{fig:setup} (a)] generating an out-of-equilibrium excitation of the free Fermi-Hubbard model [Fig.~\ref{fig:setup} (b)]. We scan over $V$ in steps of $0.1J$ in the interval $V\in[0.1,2]J$ and over $U$ in steps of $0.3J$ in the interval $U\in[0.1,4]J$. (Here and in the following we use the hopping parameter $J$ as characteristic energy scale and $1/J$ as characteristic time scale.) Since in the spin-orbital representation the entire information on $\Delta_{123}$  is contained in the spin block $\Delta_{123}^{\uparrow\uparrow\downarrow}$, we focus on the magnitude of $\Delta_{12}^{\uparrow\downarrow}$, $\Delta_{12}^{\uparrow\uparrow}$ and $\Delta_{123}^{\uparrow\uparrow\downarrow}$, as measured by the Frobenius norm (Schatten 2-norm)
\begin{equation}
	||M|| = \left(\text{Tr}M^\dagger M\right)^{1/2}.
	\label{eq:2-norm}
\end{equation}
The Frobenius norm provides an upper bound of the largest eigenvalue of $M$. The square of the Frobenius norm has been used in previous time-dependent studies as a size-extensive measure of correlations \cite{skolnik_cumulant_2013}. \\
In Fig.~\ref{fig:cumu_dyn_examp} we show a typical example for the non-equilibrium dynamics of cumulants at $V=1J$ and $U=3.1J$. 
All cumulants start with non-zero values $\Delta_{12}(t=0)$ and $\Delta_{123}(t=0)$ of the initial out-of-equilibrium state. They significantly increase immediately following the potential quench signifying the build-up of dynamical correlations in non-equilibrium dynamics. Direct visual inspection reveals that the variations of $\Delta_{12}^{\uparrow\downarrow}$, $\Delta_{12}^{\uparrow\uparrow}$ and $\Delta_{123}^{\uparrow\uparrow\downarrow}$ are correlated in time with each other. To quantify this time correlations, we calculate the equal-time limit $C_{fg}(\tau=0)$ of the normalized cross-correlation function 
\begin{equation}
	C_{f,g}(\tau) = \frac{\frac{1}{T}\int_{t_0}^T dt [f(t+\tau)-\bar f][g(t)-\bar g]}{\sigma_f\sigma_g},
	\label{eq:pearson}
\end{equation}
with $T$ the total time interval considered, the standard deviation
\begin{equation}
	\sigma_f = \sqrt{\frac{1}{T}\int_{t_0}^T dt [f(t)-\bar f]^2},
	\label{eq:std}
\end{equation}
and the mean $\bar f=\frac{1}{T}\int_{t_0}^T\, dtf(t)$ (similarly for $g$). $C_{fg}(\tau=0)$ is also referred to as the Pearson correlation coefficient \cite{pearson_1896}. With this normalization $-1\leq C_{fg}(\tau=0)\leq 1$ where $C_{fg}(\tau=0)=1(-1)$ corresponds to perfect (anti-)correlation and $C_{fg}(\tau=0)=0$ to absence of correlation in time.\\
The behavior of $C_{fg}(\tau=0)$ for different cumulant pairs in the $U$-$V$ plane is displayed in Fig.~\ref{fig:pearson_m6} and Fig.~\ref{fig:pearson_m8} for different system sizes (Fig.~\ref{fig:pearson_m6} for $M_s=6$ sites, Fig.~\ref{fig:pearson_m8} for $M_s=8$ sites). For the whole parameter scan we use $T=50 J^{-1}$, and $t_0 =10 J^{-1}$. We use a finite $t_0$ in Eq.~\ref{eq:pearson} (instead of evaluating the correlation starting with $t=0$) because the initial rise of the cumulants is always correlated and its inclusion could lead to an overestimate of the correlation coefficient. We focus here on the long-time average over the fluctuations after the initial build-up. We use $t_0=10 J^{-1}$ found to be large enough to separate the initial build-up from the fluctuations around the mean for most parameters in the $U$-$V$ plane. 
To delimit and identify structures in the $U$-$V$ landscape we also display the distribution of two characteristic variables in the $U$-$V$ plane. One is the ratio of the time-averaged correlation energy $\bar{E}_\text{cor}$ to the initial degree of excitation parameterized by $E_\text{pot}(0)$ [Figs.~\ref{fig:pearson_m6} (c) and \ref{fig:pearson_m8} (c)]. The latter is given by
\begin{equation}
	E_\text{pot}(0) = \text{Tr}D_1(t=0)V_1(t=0),
\end{equation}
while 
\begin{equation}
	E_\text{cor}(t) = \text{Tr}_{12}W_{12}\Delta_{12}(t),
	\label{eq:ecor}
\end{equation}
which in case of the Fermi-Hubbard model reduces to 
\begin{equation}
	E_\text{cor}(t) = U \sum_j\Delta_{j\uparrow j\downarrow}^{j\uparrow j\downarrow}(t).
	\label{eq:ecor_fh}
\end{equation}
The other variable measures the build-up of dynamical three-particle correlations during time evolution relative to the three-particle correlations already present in the initial state at $t=0$ prior to the quench [Figs.~\ref{fig:pearson_m6}, \ref{fig:pearson_m8} (d)],
\begin{equation}
	\delta \Delta_{123}^{\uparrow\uparrow\downarrow} = \frac{1}{T}\int_0^Tdt\, ||\Delta_{123}^{\uparrow\uparrow\downarrow}(t)|| - ||\Delta_{123}^{\uparrow\uparrow\downarrow}(0)||.
	\label{eq:cumu3_build}
\end{equation}
The contour line $\delta \Delta_{123}^{\uparrow\uparrow\downarrow}=0.65$ is also denoted in Fig.~\ref{fig:pearson_m6} and \ref{fig:pearson_m8} (a) and (b) marking quite accurately the borderline between strong and weak time correlation (or anti-correlation) between $\Delta_{123}$ and $\Delta_{12}$. We also display the borderline between high and low relative correlation energy by plotting the contour line $\bar E_\text{cor} = -0.1 E_\text{pot}(0)$ in Figs.~\ref{fig:pearson_m6} and \ref{fig:pearson_m8} (a), (b), and (c) which accurately delimits the region of strong time correlation (i.e.~$C_{\Delta_{12},\Delta_{123}}\lesssim 1$) in the cumulant dynamics. \\
Obviously, distinct parameter regimes exist for which $\Delta_{123}(t)$ and $\Delta_{12}(t)$ are strongly correlated with each other: one region pertains to small $U$ ($U\lesssim0.1J$) and a wide range of excitation energies ($0\leq V \lesssim 2J$). In this region, the cumulants build up over the whole time interval investigated of $T=50 J^{-1}$ and have not reached saturation for most $V$. This build-up is naturally strongly correlated over the whole time interval.
The other region of positive correlations can be associated with negative relative correlation energies $\bar E_\text{cor} \lesssim -0.1 E_\text{pot}(0)$ present for the whole interval of $U$ tested and moderate levels of excitation ($V\lesssim J$). Furthermore, we find for the larger system ($M_s=8$) also a region of time-correlation between $\Delta_{12}(t)$ and $\Delta_{123}(t)$ for positive correlation energy of $\bar E_\text{cor} \gtrsim 0.05 E_\text{pot}(0)$. In other regions the dynamics of $\Delta_{12}(t)$ and $\Delta_{123}(t)$ is either uncorrelated or even anticorrelated.\\
In view of the quadratic dependence of the approximate reconstruction functionals of $\Delta_{123}(t)$ on $\Delta_{12}(t)$ (Eqs.~\ref{eq:cumu3_ny}, \ref{eq:cumu3_ts}, \ref{eq:3rdm_m}) the time-correlation maps (Figs.~\ref{fig:pearson_m6} and \ref{fig:pearson_m8}) determined here from exact calculations, allow predictions for the anticipated accuracy of the TD2RDM theory. The time evolution of the many-body system should be captured quite well with the present set of reconstruction functionals in those parameter regions in the $U$-$V$ plane where the time-correlation between $\Delta_{12}(t)$ and $\Delta_{123}(t)$ is strong. As will be shown below, the approximate reconstruction functionals are reasonably accurate as long as the build-up of three-particle correlations over time (Eq.~\ref{eq:cumu3_build}) remains moderate.\\
To assess the accuracy of the reconstruction functionals locally in time and without the accumulation of errors during time evolution, we also perform exact calculations of both $D_{12}(t)$ and $D_{123}(t)$ and compare the latter with the reconstructed $D_{123}^\text{R}(t)$ using the exact $D_{12}(t)$ as input for the reconstruction,
\begin{equation}
	\delta[D_{123}(t)] = ||D_{123}^{\rm R}[D_{12}^{\rm exact}(t)]-D_{123}^{\rm exact}(t)||.
	\label{eq:err_3rdm}
\end{equation}
Taking into account the cumulant expansion of $D_{123}$ (see Eq.~\ref{eq:3rdm_cumu}) this error coincides with the error in the three-particle cumulant $\delta[D_{123}] = \delta[\Delta_{123}]$ as only the latter is subject to reconstruction errors. In Fig.~7 we present exemplary results for $\delta[D_{123}(t)]$ for the $\Delta_{123}^{\uparrow\uparrow\downarrow}$ block and for the parameters $V=0.8J$ and $U=1.9J$ localized in the region of strong temporal correlations [Fig.~\ref{fig:pearson_m6} (a), (b)] as well as moderate build-up of three-particle correlations over time [Fig.~\ref{fig:pearson_m6} (d)]. As the bare Valdemoro approximation neglects $\Delta_{123}$ entirely, its error is largest and corresponds to the exact value of $\Delta_{123}$ itself. The bare NY, M, and TS perform better with the NY-reconstruction performing best.
\begin{figure}[t]
	\includegraphics[width=\columnwidth]{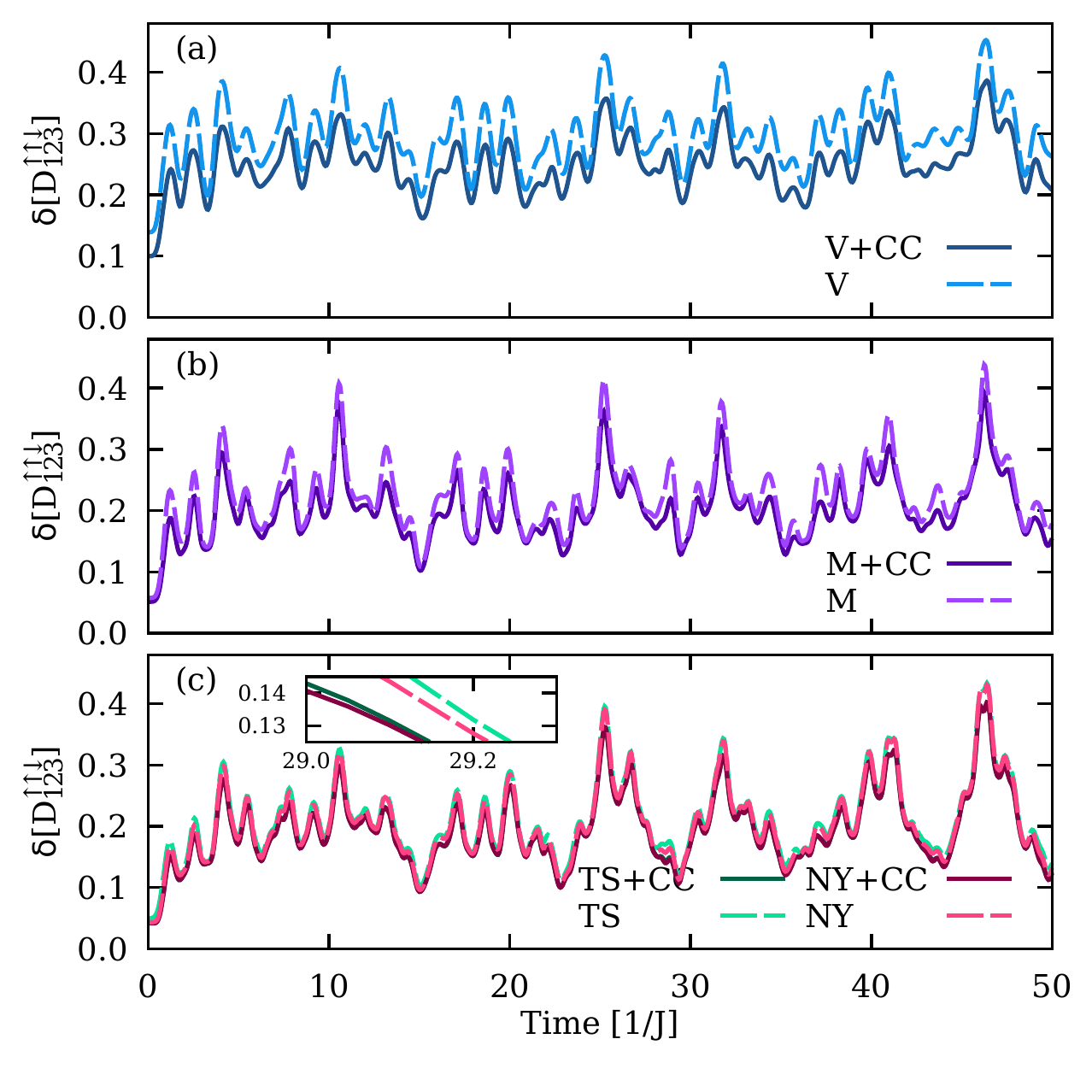}
	\caption{Six-site Fermi-Hubbard model at half filling ($M_s=6$) starting from the many-body ground state in the potential with $V=0.8 J$ defining the out-of-equilibrium initial state in the potential-free Fermi-Hubbard model after the quench. The interaction parameter is $U=1.9J$. Error in the three-particle reconstruction $\delta[D_{123}(t)]$ (Eq.~\ref{eq:err_3rdm}) for different reconstruction functionals (a) Valdemoro (V) (Eq.~\ref{eq:3rdm_v}) with and without contraction consistency (CC), (b) Mazziotti (M) (Eq.~\ref{eq:3rdm_m}) with and without CC, and (c) Tohyama-Schuck (TS) (Eq.~\ref{eq:3rdm_ts}) and Nakatsuji-Yasuda (NY) (Eq.~\ref{eq:3rdm_ny}) with and without CC. The inset in (c) shows a zoom into a region where the difference between TS+CC and NY+CC becomes visible. 
	}
	\label{fig:qor_3rdm}
\end{figure}
The difference between NY and TS is very small indicating that the normalization $\mathcal{N}$ does not play a significant role in this case. Inclusion of the CC corrections improves the performance of all reconstruction functionals (Fig.~\ref{fig:qor_3rdm}). As expected, the changes are largest for V+CC for which the CC correction given by the orthogonal component of the cumulant, $\Delta_{123;\perp}$ (Eq.~\ref{eq:cumu_corr}), represents the only contribution to $\Delta_{123}$. For the TS and NY functionals, on the other hand, the corrections due to CC are small in this particular case.\\
To further probe the accuracy of the reconstruction functionals within the TD2RDM theory locally in time in more detail we now take into account that only a fraction of the elements of the full 3RDM enters the equations of motion of the 2RDM via the collision operator $C[D_{123}]$ (see Eq.~\ref{eq:eom}). We therefore determine the corresponding relative error in the collision operator
\begin{align}
	\delta[C(t)] &= \frac{||C[D_{123}^{\rm R}(t)]-C_{\rm exact}(t)]||}{||C_{\rm exact}(t)||},
	\label{eq:err_cop}
\end{align}
using the exact input from $D_{12}^\text{exact}$ in $D_{123}^\text{R}$.
\begin{figure}[t]
	\includegraphics[width=\columnwidth]{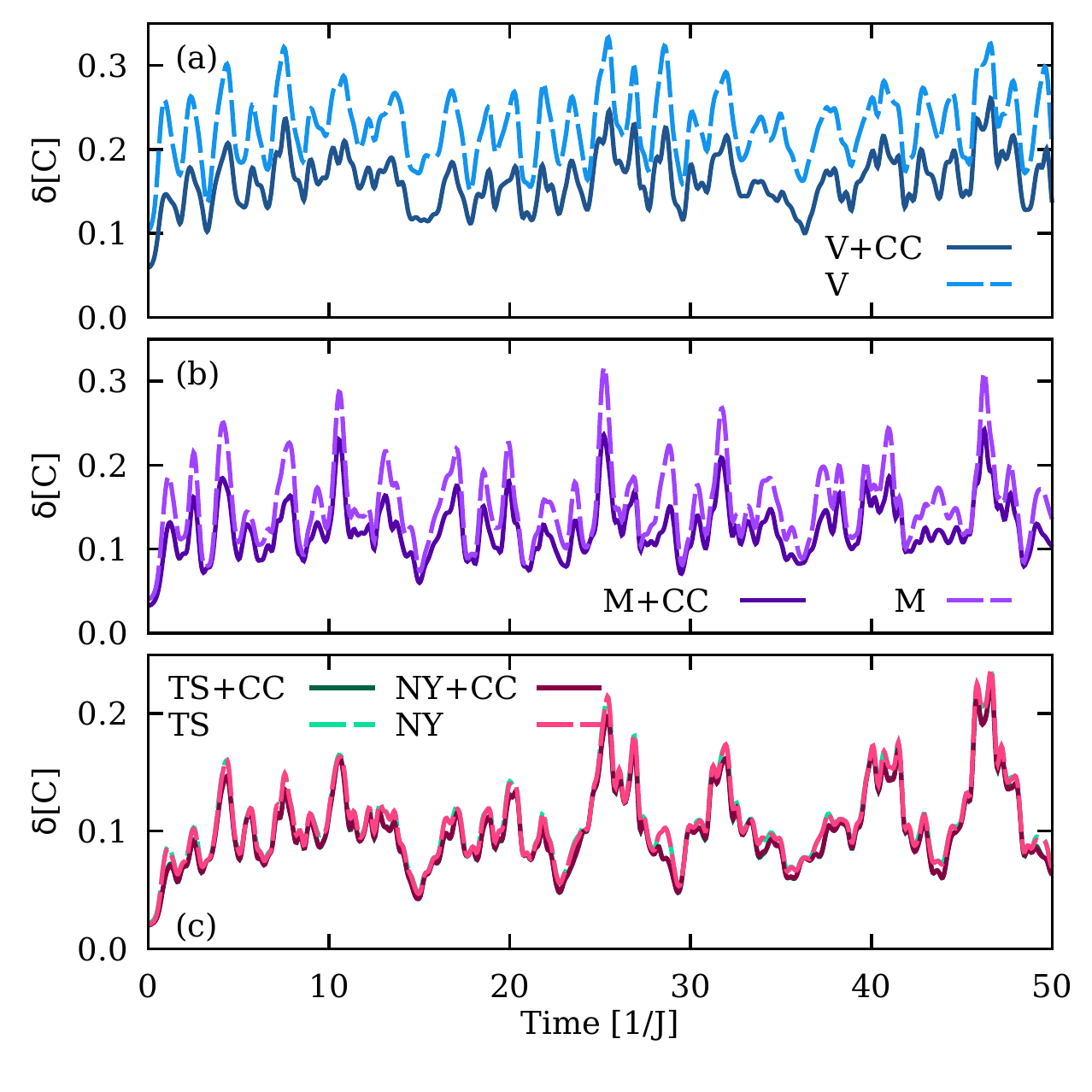}
	\caption{As in Fig.~\ref{fig:qor_3rdm} however for the relative error of the collision operator $\delta[C]$ (Eq.~\ref{eq:err_cop}). Note the different scale of the vertical axis in (c) as compared to (a) and (b).
	}
	\label{fig:qor_cop}
\end{figure}
The error in the collision operator (shown in Fig.~\ref{fig:qor_cop} for the $\uparrow\uparrow\downarrow$-block) mirrors closely that of $\Delta_{123}$ (Fig.~\ref{fig:qor_3rdm}). It is largest for the V functional and smallest for the NY+CC functional. In the following benchmark calculations of the non-equilibrium dynamics of the Fermi-Hubbard model for different pairs of ($U$, $V$) we will restrict ourselves to these two functionals which provide a clear indication of the bandwidth of the expected accuracy.\\
It is furthermore instructive to directly compare the time-local reconstruction error $\delta[D_{123}(t)]$ (Eq.~\ref{eq:err_3rdm}) of the V+CC and NY+CC reconstruction functionals for the cumulants with the norm of the cumulants themselves (Fig.~\ref{fig:qor_cum}).
\begin{figure}[t]
	\includegraphics[width=\columnwidth]{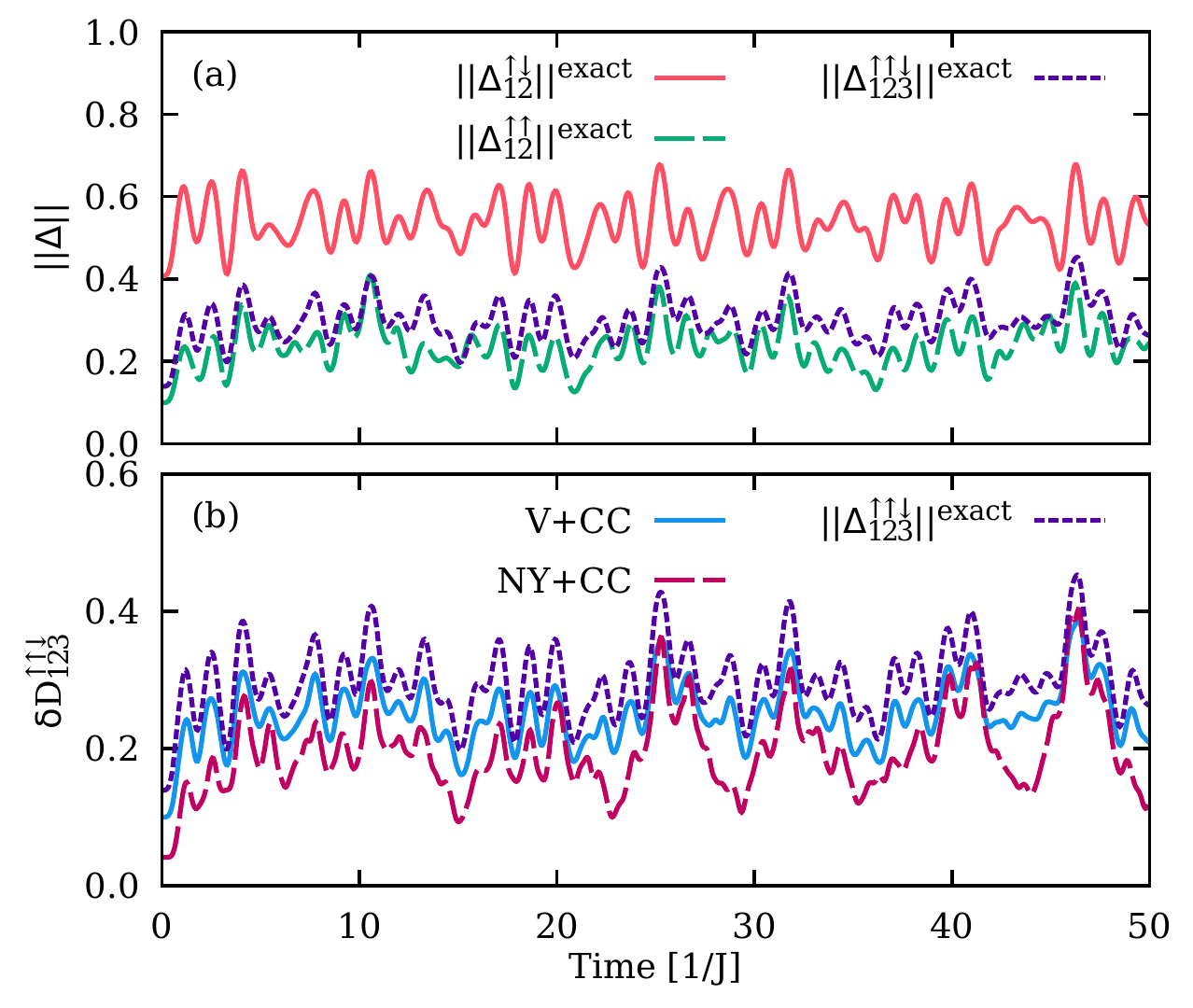}
	\caption{Same system as in Fig.~\ref{fig:qor_3rdm}. (a) Dynamics of the two-particle ($\Delta_{12}^{\uparrow\uparrow}$ and $\Delta_{12}^{\uparrow\downarrow}$) and three-particle ($\Delta_{123}^{\uparrow\uparrow\downarrow}$) cumulants as measured by the Frobenius norm and obtained from an exact wavefunction calculation. (b) Time-resolved error of the reconstruction $\delta[D_{123}^{\uparrow\uparrow\downarrow}]$ using the NY+CC and V+CC functionals compared with $||\Delta_{123}^{\uparrow\uparrow\downarrow}||^\text{exact}$ [same curve as in (a)].
	}
	\label{fig:qor_cum}
\end{figure}
Note that the norm $||\Delta_{123}(t)||$ [Fig.~\ref{fig:qor_3rdm} (a)] coincides with the error of the bare V reconstruction functional in which the three-particle is neglected. In turn, the difference to the V+CC functional [Fig.~\ref{fig:qor_cum} (b)] directly indicates the size of $\Delta_{123;\perp}$ included by enforcing contraction consistency. This correction amounts in the present system to an approximate scaling factor of $\approx 0.85\pm 0.05$. The time-local reconstruction functional NY+CC improves the reconstruction substantially compared to V+CC. We find that NY+CC performs better in regions where $\Delta_{123}^{\uparrow\uparrow\downarrow}$ has local minima but performs similarly as the V+CC in regions of local maxima. This gives an indication of current limitations of the reconstruction accuracy and also useful hints for directions of future improvements.\\
We now analyze the time-averaged reconstruction error in the collision operator (Eq.~\ref{eq:err_cop}) in the $U$-$V$ plane (Fig.~\ref{fig:qor_scan_cop}).
\begin{figure}[t]
	\includegraphics[width=\columnwidth]{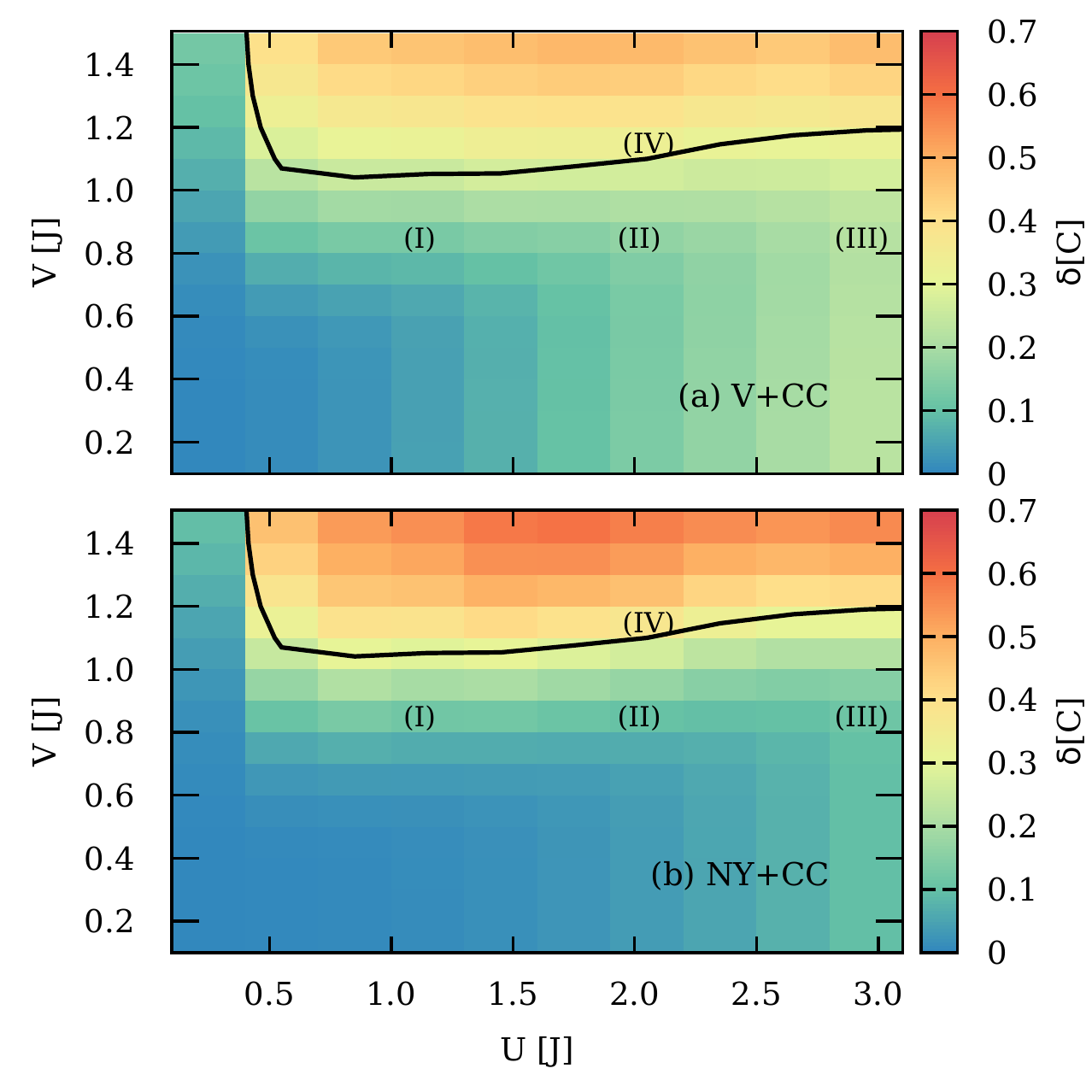}
	\caption{Six-site Fermi-Hubbard model at half filling ($M_s=6$) and varying pair interaction $U$ initially ($t=0$) confined by the harmonic potential of strength $V$ and released for $t>0$. Shown is the reconstruction error in the $U$-$V$ plane measured by the time-averaged relative error of the collision operator $\delta[C]$ (Eq.~\ref{eq:err_cop}) using (a) the V+CC reconstruction functional and (b) the NY+CC reconstruction functional. For the points in the $U$-$V$ plane marked by (I)-(IV) the time evolution of the occupation number $n_1(t)$ is displayed in Fig.~\ref{fig:scan_n1} and their error in Fig.~\ref{fig:scan_occ}. The black solid line corresponds to the same contour line as in Fig.~\ref{fig:pearson_m6}.
	}
	\label{fig:qor_scan_cop}
\end{figure}
The time-averaged error closely mirrors the behavior of the equal-time cross-correlation between the two-particle and three-particle cumulants (Eq.~\ref{eq:pearson}, Fig.~\ref{fig:pearson_m6}). For a Fermi-Hubbard system with weak inter-particle interactions $U\lesssim0.1J$ the reconstruction error in the collision operator is very small for both the V+CC and the NY+CC reconstruction. For much larger $U$ of up to $U\approx3J$ and moderately strong initial excitation ($V\lesssim J$) the NY+CC reconstruction performs markedly better. Remarkably, this region is quite faithfully delimited by the region where the build-up over time of the correlations (Eq.~\ref{eq:cumu3_build}) is moderate, $\delta\Delta_{123}^{\uparrow\uparrow\downarrow}\lesssim 0.65$. Nevertheless, it should be pointed out that the accuracy also of this reconstruction is limited for larger $U\gtrsim 3J$. Interestingly, in the $U$-$V$ region of time-anti-correlated or uncorrelated dynamics of the cumulants (Figs.~\ref{fig:pearson_m6} and \ref{fig:pearson_m8}) the time-local reconstruction within NY+CC can cause even larger errors than the V+CC. This is not surprising in view of the fact that the V+CC approximation neglects (apart from the CC correction) $\Delta_{123}$ entirely and does not enforce time correlations as the NY+CC reconstruction does through the quadratic dependence of $\Delta_{123}^\text{NY}$ on $\Delta_{12}$ (see Eq.~\ref{eq:cumu3_ny}). Also this observation may point to avenues for further improvements of reconstruction functionals.
\section{Self-consistent propagation of the 2RDM}\label{sec:res_bench}
\begin{figure}[t]
	\includegraphics[width=\columnwidth]{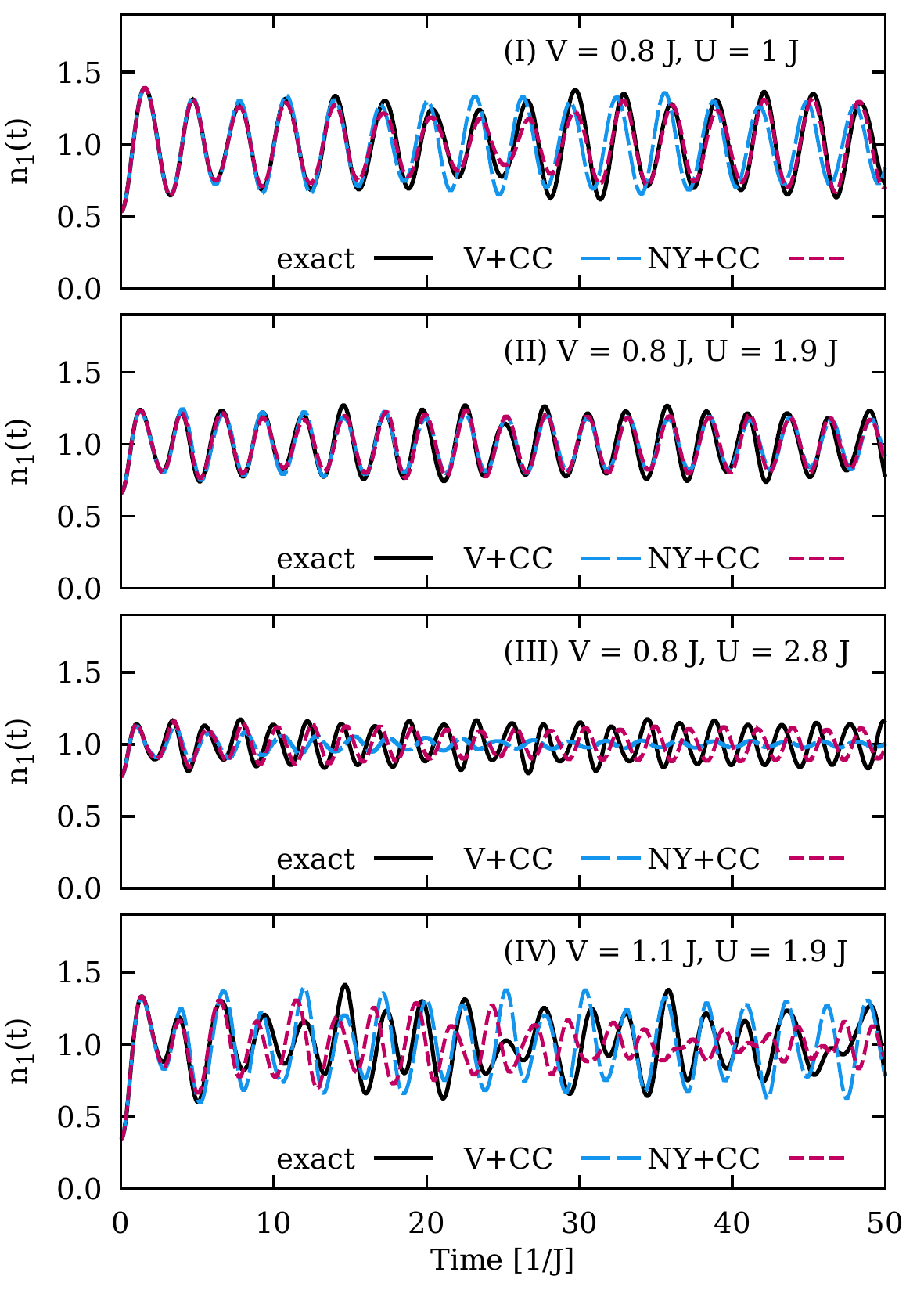}
	\caption{System as in Fig.~\ref{fig:qor_scan_cop}. Shown is the time evolution of the occupation of site $1$, $n_1(t)$ (see Fig.~\ref{fig:setup}), as predicted by TD2RDM using the V+CC and NY+CC functionals and compared with the exact results for different parameter combinations of ($U$,$V$) as indicated in the frame and marked in Fig.~\ref{fig:qor_scan_cop} . 
	}
	\label{fig:scan_n1}
\end{figure}

We present now examples of the fully self-consistent solution of the equations of motion of the 2RDM (Eq.~\ref{eq:eom}) starting from the pure excited state, the many-body ground state in the potential $V_i(t)$ prior to the quench at $t=0$. We present examples of these simulations for different parameters in the $U$-$V$ plane marked in Fig.~\ref{fig:qor_scan_cop}. As observable for the quench dynamics we chose the occupation $n_1(t) = D_1^1(t)$ of the first site. We compare the results of the TD2RDM theory for a given reconstruction functional with the corresponding exact calculations (Fig.~\ref{fig:scan_n1}). 
As a figure of merit we use the time-integrated deviation 
\begin{equation}
	\delta n_1 =  \frac{\int_0^Tdt\, |n_1^{\rm exact}(t)-n_1^{\rm R}(t)|}{\int_0^Tdt\, n_1^{\rm exact}(t)},
	\label{eq:diff_n1}
\end{equation}
sensitively probing the amplitude, frequency, and phase of the quench-induced density fluctuations (Fig.~\ref{fig:scan_n1}).
\begin{figure}[t]
	\includegraphics[width=\columnwidth]{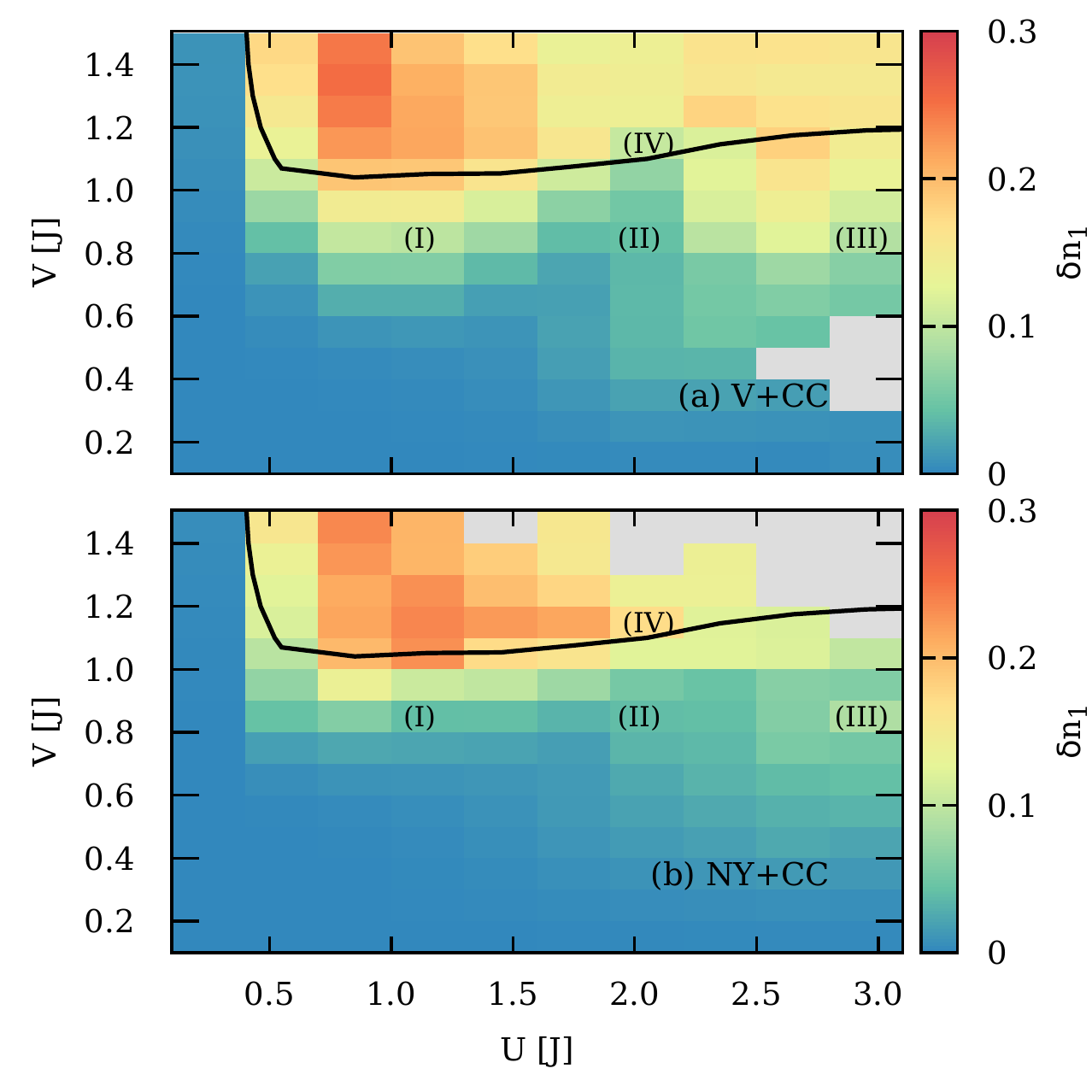}
	\caption{System as in Fig.~\ref{fig:qor_scan_cop}. Shown is the time-integrated error $\delta n_1$ (Eq.~\ref{eq:diff_n1}) of the occupation number $n_1(t)$ predicted by TD2RDM relative to the exact result in the $U$-$V$ plane. The black line denotes the $\delta \Delta_{123}^{\uparrow\uparrow\downarrow}=0.65$ contour of the build-up of three-particle correlations over time (Eq.~\ref{eq:cumu3_build}) (same as in Fig.~\ref{fig:pearson_m6}). The gray areas mark regions where the method did not meet the convergence criteria discussed in Appendix \ref{app:purify}.
	}
	\label{fig:scan_occ}
\end{figure}
We note that the convergent and accurate propagation of the 2RDM requires, in addition to an accurate three-particle cumulant reconstruction functional, also the preservation of $N$-representability which is a priori not guaranteed when errors due to the approximate reconstruction functionals pile up. $N$-representability is approximately restored during propagation by purification ``on the fly" (see \cite{lackner_propagating_2015, lackner_high-harmonic_2017}). The specific purification algorithm employed in the present simulation is summarized in Appendix \ref{app:purify}.\\
The time evolution of $n_1(t)$ for selected values of $U$ and $V$ marked by roman numbers in Fig.~\ref{fig:qor_scan_cop}  are displayed for $t\leq 50 J^{-1}$ in Fig.~\ref{fig:scan_n1}. For weak on-site interaction $U\lesssim 0.1J$ both the V+CC and the NY+CC reconstruction functionals yield excellent agreement with the exact results over a wide range of out-of-equilibrium excitations $0\leq V\leq 1.5J$ (see also Fig.~\ref{fig:scan_occ}). For stronger $U$ and intermediate quenches with $V$ up to $V=0.8J$ [Fig.~\ref{fig:scan_n1} (a)-(c)] we observe excellent agreement for the NY+CC reconstruction which performs better than the V+CC reconstruction. For larger $V$ [e.g.~$V=1.1J$ and $U=1.9J$, Fig.~\ref{fig:scan_n1} (d)] deviations for both functionals from the exact result are larger with the V+CC reconstruction functional performing slightly better.\\
To survey the accuracy of the site occupation $n_1(t)$ we display in Fig.\ref{fig:scan_occ} the time-integrated deviations $\delta n_1$ from the exact result (Eq.~\ref{eq:diff_n1}) for the $6$-site Fermi-Hubbard model in the $U$-$V$ plane. This distribution closely resembles the equal-time correlation between $\Delta_{12}$ and $\Delta_{123}$ (see Fig.~\ref{fig:pearson_m6}, Eq.~\ref{eq:pearson}). Obviously, the limitation to moderate build-up of three-particle correlations over time ($\delta \Delta_{123}^{\uparrow\uparrow\downarrow}\lesssim 0.65$) is one reliable predictor for accurate long-term simulations of the correlated non-equilibrium dynamics. We emphasize that the present figure of merit ($\delta n_1\lesssim 0.1$, i.e.~the area shaded green or blue in Fig.~\ref{fig:qor_scan_cop}) puts the theory to a fairly stringent test. Even when $\delta n_1 \gtrsim 0.3$ the agreement with the exact calculation is qualitatively and even semi-quantitatively satisfactory, capturing key features of the fluctuations even though not in all details (see, e.g., Fig.~\ref{fig:scan_n1} IV). Likewise, when the purification protocol does not fully converge relative to the criteria imposed (see Appendix \ref{app:purify}), the results for the time-dependent occupation numbers still contain qualitatively correct information on the mean occupation and dominant frequencies.\\
We now turn to larger systems ($M_s=18, 20$) and time scales for which exact or highly accurate wavefunction based methods (such as MPS) are presently still a challenge. We demonstrate the straightforward applicability of the TD2RDM theory for such systems. For a meaningful comparison with mean-field methods such as TDHF we restrict ourselves to a weakly correlated Fermi-Hubbard model with $U=0.1 J$ but a high degree of excitation. We start with an initial state where all $M_s/2$ sites around the center of the system are doubly occupied amounting to a quench in the limit $V\rightarrow \infty$. For the system size $M_s=18$ [Fig.~\ref{fig:large_site_1} (a)] we can still compare with exact propagation for the time interval $0\leq t \leq 80 J^{-1}$. 
\begin{figure}[t]
	\includegraphics[width=\columnwidth]{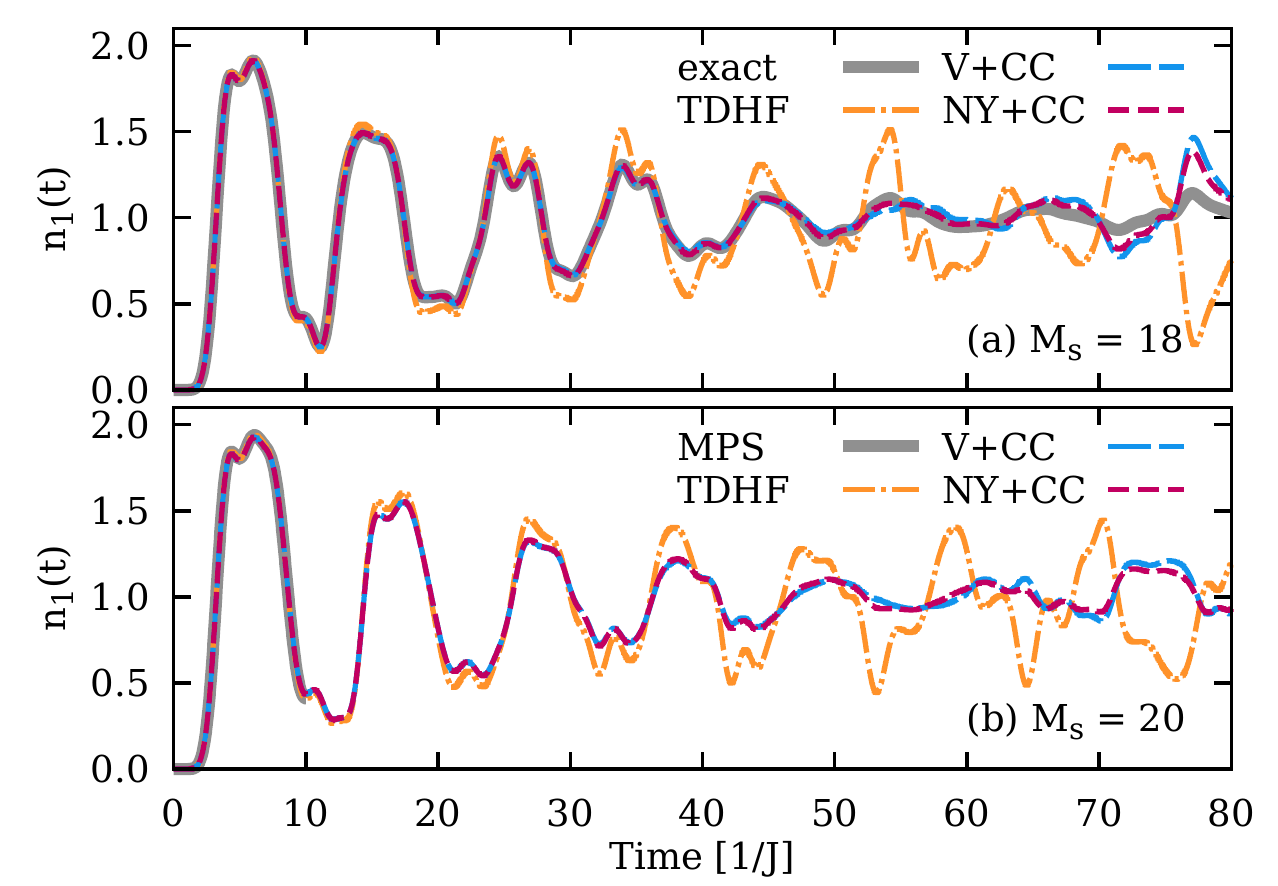}
	\caption{The Fermi-Hubbard model with (a) $M_s=18$ sites and (b) $M_s=20$ sites at half filling and $U=0.1 J$. All particles are initially located at the center of the system amounting to a potential quench of $V\rightarrow\infty$. The exact solution of the Schr\"odinger equation in (a) has been obtained using a Trotter decomposition of the sparse time-evolution operator (time step $dt=0.025 J^{-1}$), and via the one-site time-dependent variational principle with matrix product state bond dimension of $\chi=512$ ($dt=0.025 J^{-1}$) in (b). Convergence with bond-dimension and time step has been verified to be within an absolute precision of $10^{-3}$ for the expectation values shown. We compare with the TD2RDM prediction using the V+CC and NY+CC reconstruction functionals as well as to the mean-field solution within time-dependent Hartree-Fock (TDHF) theory.}
	\label{fig:large_site_1}
\end{figure}
For $M_s=20$ [Fig.~\ref{fig:large_site_1} (b)] we can compare to MPS calculations. The latter are, however, limited to short times $t\lesssim 10J^{-1}$. For $M_s=18$ we find excellent agreement with the exact results up to $t\approx 60 J^{-1}$ and still reasonable agreement for longer times. For $M_s=20$ we find excellent agreement with the MPS results for the short time interval for which the MPS data could be generated. We emphasize that increasing the system sizes here from $M_s=18$ to $M_s=20$ does not pose any major challenge for TD2RDM theory. The extension towards larger systems approaching extended periodic systems thus appears feasible. We also present in Fig.~\ref{fig:large_site_1} a comparison with a time-dependent Hartree-Fock (TDHF) simulation as a representative mean-field description within which larger systems are accessible. Even though the system is only weakly correlated, TDHF fails after a short time interval ($t\gtrsim 35 J^{-1}$) and substantially overestimates the oscillation amplitude of $n_1(t)$ (Fig.~\ref{fig:large_site_1}). By contrast the TD2RDM method is able to correctly capture the dynamics in this system for extended periods of time.
\section{Conclusions and Outlook}\label{sec:concl}
In this paper, we have applied the time-dependent two-particle reduced density matrix (TD2RDM) theory to the non-equilibrium dynamics of the finite-size Fermi-Hubbard model at half-filling in 1D for a wide range of number of sites and interaction strengths $U$ and initial out-of-equilibrium excitations controlled by the potential strength $V$ of the quench. The Fermi-Hubbard model serves here as a benchmark model to demonstrate the applicability and performance of the theory to extended systems with non-negligible correlations relevant for current research in condensed matter physics and ultra-cold atoms.\\
The TD2RDM theory fully incorporates two-particle correlations and includes approximate three-particle correlations via reconstruction functionals. Key to an accurate description of the dynamics within the TD2RDM theory is the reconstruction of the 3RDM, $D_{123}$, by means of the 2RDM, $D_{12}$, to close the equations of motion, and application of contraction consistency. The underlying assumption of the closure is the existence of a sufficiently accurate reconstruction functional of $\Delta_{123}$. Currently used functionals assume $\Delta_{123}$ to be local in time, i.e.~they do not take into account possible memory effects. The existence of such a reconstruction functional is guaranteed for the ground state via Rosina's theorem \cite{rosina_book_1968, mazziotti_book_2007} but its extension to time-dependent settings is currently unknown. By comparing with exact results for small system sizes we have analyzed the dynamics of both two- and three-particle cumulants. Over a wide range of $U$ and $V$ we could identify parameter regimes in which the dynamics of the three-particle and two-particle cumulants are indeed strongly correlated in time with each other, a key prerequisite for the applicability of current state-of-the art time-local reconstruction functionals. 
For this particular model system we could show the applicability and accuracy of the TD2RDM theory well into the regime of moderately strong correlations ($U\lesssim 3J$), of moderately strong out-of-equilibrium excitations ($V\lesssim J$), and for long propagation times (close to hundred time units $J^{-1}$).  As an approximate parameter controlling the applicability of TD2RDM theory with the present functionals we could identify the difference $\delta \Delta_{123}$ between the dynamically built up and the initially present (ground state) three-particle correlations. The present observation of the key role of temporal correlations between the two- and three-particle cumulants as well as of the build-up of three-particle correlations over time will provide us with directions for further improvements of the reconstruction functionals. Moreover, they may serve as a guidance for the applicability of TD2RDM theory for systems where exact benchmarks are not available.\\
Application to larger systems indicates that TD2RDM theory is still capable of providing accurate results and may outperform wavefunction based methods. We have showcased an example in the regime of weak interactions and a high degree of excitation in a system with $18$ sites, where a numerically exact solution of the full Schr\"odinger equation is still possible, and with $20$ sites where a MPS solution can be generated, however only for a limited time span ($\leq 10 J^{-1}$). Based on the excellent agreement with exact results we conclude that the TD2RDM theory has the potential to develop into a versatile tool to study the correlated and strongly driven dynamics of extended models relevant to ultracold atoms and solid state physics. Extensions of the TD2RDM theory to extended systems in two- and three-dimensional systems are planned.
\section*{Acknowledgements}
We thank Daniel Wieser for helpful discussions. IB thanks the Simons Foundation for the great hospitality and support during her research visit at the CCQ of the Flatiron Institute.  The Flatiron Institute is a division of the Simons Foundation. We acknowledge support from the Max Planck-New York City Center for Non-Equilibrium Quantum Phenomena, 
Cluster of Excellence `CUI: Advanced Imaging of Matter'- EXC 2056 - project ID. This research was funded by the WWTF grant MA-14002, the Austrian Science Fund (FWF) grant P 35539-N, the FWF doctoral college Solids4Fun, as well as the International Max Planck Research School of Advanced Photon Science (IMPRS-APS). Calculations were performed on the Vienna Scientific Cluster (VSC4). 
\appendix
\section{Equations of motion for the 2RDM}\label{app:eom}
We present here the explicit expression for the equation of motion for $D_{j_1\uparrow j_2\downarrow}^{i_1\uparrow i_2\downarrow}$ in the basis of spin ($\sigma=\uparrow,\downarrow$) orbitals localized at site $i$. Inserting the reconstructed $D_{j_1\uparrow j_2\uparrow j_3\downarrow}^{i_1\uparrow i_2\uparrow i_3\downarrow}$ (Eq.~\ref{eq:D123_rec}) into Eq.~\ref{eq:eom} together with Eq.~\ref{eq:h_1} and Eq.~\ref{eq:h_int} yields
\begin{align}
	i\partial_t D_{j_1\uparrow j_2\downarrow}^{i_1\uparrow i_2\downarrow} &= 
	 \sum_n h_{j_1}^n D_{n\uparrow j_2\downarrow}^{i_1\uparrow i_2\downarrow} 
	+\sum_n  h_{j_2}^n D_{j_1\uparrow n\downarrow}^{i_1\uparrow i_2\downarrow} \nonumber \\
	&+ U \delta_{j_1,j_2}D_{j_1\uparrow j_2\downarrow}^{i_1\uparrow i_2\downarrow} \nonumber \\
	&- \sum_n h_{n}^{i_1} D_{j_1\uparrow j_2\downarrow}^{n\uparrow i_2\downarrow} 
	- \sum_n h_{n}^{i_2} D_{j_1\uparrow j_2\downarrow}^{i_1\uparrow n\downarrow} \nonumber \\
	&- U \delta^{i_1,i_2}D_{j_1\uparrow j_2\downarrow}^{i_1\uparrow i_2\downarrow} \nonumber \\
	&+ U D_{j_1\uparrow j_2\uparrow j_1\downarrow}^{j_1\uparrow i_2\uparrow i_1\downarrow}
	+ U D_{j_1\uparrow j_2\uparrow j_2\downarrow}^{i_1\uparrow j_2\uparrow i_2\downarrow}\nonumber \\
	&- U D_{i_1\uparrow j_2\uparrow j_1\downarrow}^{i_1\uparrow i_2\uparrow i_1\downarrow}
	- U D_{j_1\uparrow i_2\uparrow j_2\downarrow}^{i_1\uparrow i_2\uparrow i_2\downarrow},
\end{align}
with
\begin{equation}
	h_j^i = -J\delta_{j}^{i+1} - J \delta_{j}^{i-1}.
\end{equation}
%
\section{Mazziotti reconstruction functional}\label{app:recon}
The Mazziotti reconstruction functional \cite{mazziotti_pursuit_1999} is given in the basis of natural orbitals by
\begin{equation}
	\Delta_{j_1j_2j_3}^{i_1i_2i_3\ {\rm M}} = -\frac{1}{\chi_{j_1j_2j_3}^{i_1i_2i_3}-3}\,
	\hat A\sum_n \Delta_{j_1j_2}^{i_1n}\Delta_{nj_3}^{i_2i_3}
	\label{eq:cum3_m}
\end{equation}
with 
\begin{equation}
	\chi_{j_1j_2j_3}^{i_1i_2i_3} = \nu_{j_1} +\nu_{j_2}+\nu_{j_3}+\nu_{i_1}+\nu_{i_2}+\nu_{i_3},
\end{equation}
and $\nu_k$ the eigenvalues of the 1RDM (i.e.~the natural occupation numbers). For propagating the equations of motion one in a single-particle basis one has to perform a basis transformation. Elements of the type $\Delta_{xxx}^{ooo}$ and $\Delta_{oou}^{ouu}$, where $o$ denotes an occupied and $u$ an unoccupied orbital, remain undetermined due to the divergence of the denominator in Eq.~\ref{eq:cum3_m} and are chosen to be zero, as suggested in \cite{mazziotti_pursuit_1999}. 
\section{Purification}\label{app:purify}
The error through reconstruction of the 3RDM in the equation of motion (Eq.~\ref{eq:eom}) typically accumulates over time such that the propagation becomes unstable \cite{akbari_challenges_2012}. Similar instabilities are found in NEGF methods \cite{joost_dynamically_2022}. We have recently shown \cite{lackner_propagating_2015} that these instabilities can be prevented by enforcing a subset of necessary N-representability conditions during the propagation. N-representability refers to the necessary and sufficient conditions a RDM has to fulfill to represent a proper reduction of a fermionic many-body wavefunction (or many-body density matrix if ensemble N-representability is concerned) \cite{garrod_reduction_1964}. While for the 1RDM it is sufficient that its eigenvalues lie within the interval $\nu_i\in[0,1]$ for ensemble N-representability \cite{coleman_structure_1963, parr_density_1989}, the pure state N-representability problem leads to so-called generalized Pauli constraints \cite{klyachko_quantum_2006, altunbulak_pauli_2008, schilling_pinning_2013, schilling_generalized_2018}. For the 2RDM
constructive methods exist to obtain a set of necessary ensemble N-representability conditions (see e.g.~\cite{garrod_reduction_1964, coleman_structure_1963, mazziotti_structure_2012}), but only a limited number of these conditions can be enforced in numerical computations, especially in a time-dependent setting. The problem of sufficient conditions of pure-state N-representability is still widely open \cite{ayers_necessary_2006}.\\
Within the TD2RDM theory, we have shown that enforcing the positive semi-definiteness of the two-particle RDM (D-condition) and the corresponding condition on the two-hole RDM,
\begin{equation}
	Q_{12} = \hat A I_1 I_2 -\hat A D_1 I_2 + D_{12},
\end{equation}
(the Q-condition) is sufficient to stabilize the propagation. Moreover, enforcing the D- and Q-condition substantially improves the accuracy of all physical observables, even when the equations of motion remain stable. A similar stabilizing effect has been observed within the G1-G2 scheme of NEGF methods \cite{joost_dynamically_2022}. The additional G-condition, positive-semidefiniteness of the particle-hole RDM, was empirically found to be fulfilled when the D- and Q-conditions are. 
\\
The successful purification enforces positive-semidefiniteness of the 2RDM and the two-hole RDM in the least invasive way which implies preserving their diagonal and off-diagonal traces, as well as preserving energy after purification. Our purification protocol utilizes the unitary decomposition of the 2RDM \cite{lackner_high-harmonic_2017}. To this end we determine the component of the 2RDM with negative eigenvalues (i.e.~geminal occupation numbers) $\eta_i$
\begin{equation}
	D_{12}^{<} = \sum_{\eta_i<0} \eta_i|g_i\rangle \langle g_i|,
	\label{eq:2rdm_def}
\end{equation}
and, analogously, the corresponding defective part of the two-hole $Q_{12}^{<}$. Subtracting these defective parts from $D_{12}$ and $Q_{12}$ would restore a positive semi-definite matrix. However, such a procedure without constraints would violate conservation of $D_1$ as well as of the energy. Therefore, we have to enforce in addition 
\begin{equation}
	\text{Tr}_2D_{12}^{<} = 0,
	 \label{eq:pur_tr_0}
\end{equation}
i.e.~the subtracted part $D_{12}^{<}$ must reside in the kernel. Moreover, the (correlation) energy must be preserved
\begin{equation}
	\text{Tr}_{12}W_{12}D_{12}^{<} = 0.
	\label{eq:pur_ecor_0}
\end{equation}
The part of $D_{12}^{<}$ meeting these requirements Eqs.~\ref{eq:pur_tr_0} and \ref{eq:pur_ecor_0} is denoted by $D_{12}^{<;E}$. We thus arrive at the purification formula for the 2RDM \cite{joost_dynamically_2022}
\begin{equation}
	D_{12}' = D_{12} - D_{12;K}^{<;E} - Q_{12;K}^{<;E},
	\label{eq:purify}
\end{equation}
which we apply iteratively each time the smallest geminal occupation number $\eta_i$ drops below a threshold value until the threshold value is reached. We would like to point out that the results depend only very weakly on the threshold value as well as the maximal number of steps applied in the iterative process as long the iteration convergences and the smallest geminal occupation number is close to zero (but can still be slightly negative). Since calculating the geminal occupation numbers through exact diagonalization is numerically costly [scaling as $O(M_s^6)$ with the number of sites] we restrict ourselves to applying Eq.~\ref{eq:purify} only once each time the smallest geminal occupation number drops below zero for the largest systems in the present paper (with $M_s=18$ and $M_s=20$) to save computational time. We have checked that the obtained results are converged with respect to the time step $dt$ of the propagation (i.e.~the number of time steps within the whole time interval $[0,T]$). This also means that applying different purification schemes (with respect to the threshold  on the smallest geminal occupation number and the number of iterative steps) will give the same results on the level of accuracy set by the threshold.\\
Reaching numerical convergence as a function of the number of time steps of the propagation when purification is applied poses a challenge in case of large reconstruction errors. We have used an overall global time step $dt$, however, within each time step we use a time-adaptive propagation (using a Runge-Kutta-Fehlberg propagator of $4^{th}$ and $5^{th}$ order) to split each time step into sub-steps within which the prescribed tolerance of the local error is reached. Whenever the smallest geminal occupation number falls below a certain threshold, purification is applied after the global time step. For the scan in Fig.~\ref{fig:scan_occ} we have applied a threshold of $-10^{-4}$ and the maximal allowed number of iterations to reach this threshold is set to $100$. We observe that in regions of large errors in the reconstruction the iterative purification often does not reach the threshold (this happens mostly in the lower right corner of Fig.~\ref{fig:scan_occ}). This does not lead necessarily to instabilities. In fact, we did not observe any instabilities for all values of $U$ and $V$ scanned in Fig.~\ref{fig:scan_occ}. However, frequent applications of purification and large numbers of iterations required to reach the threshold or not reaching the threshold at all may induce undesirable numerical noise. This may prevent convergence as a function of the size of the global time step $dt$ (i.e.~as a function of the number of time steps used in the propagation) in cases when the uncontrolled noise accumulates (Fig.~\ref{fig:scan_occ} gray regions). These areas are determined by discarding all results with a local error in the physical observable of $>5\times 10^{-3}$. Even if the simulation does not meet strict convergence criteria some of the observables are reasonably well represented, e.g.~the mean (i.e.~time-averaged) occupation number.
\section*{References}
%
\end{document}